\documentclass{aastex631}
\pdfoutput=1
\usepackage{color}
\usepackage{soul}
\usepackage{graphicx}

\received{}
\revised{}
\accepted{}

\submitjournal{AAS Journals}
\shorttitle{NEID Solar}
\shortauthors{Ford et al.}

\newcommand{\mps}{m s$^{-1}$}
\newcommand{\kmps}{km s$^{-1}$}
\newcommand{\mpsphr}{m s$^{-1}$ hr$^{-1}$}
\newcommand{\mpsphrporder}{m s$^{-1}$  hr$^{-1}$ order$^{-1}$}
\newcommand{\phiesta}{FIESTA~}

\newcommand{\stdmeandailyrv}{2.11}
\newcommand{\medianrmsbinnedrv}{0.489}
\newcommand{\numobsused}{117062}

\newcommand{\numbinnedobsused}{20438}

\newcommand{\numdaysmeanrv}{698}

\newcommand{\numdaysmeanslope}{544}
\newcommand{\rvslopevstimeslope}{-0.179}
\newcommand{\rvslopevstimerms}{0.156}

\newcommand{\amplitudesinjointslopemodel}{0.0817}
\newcommand{\amplitudecosjointslopemodel}{-0.0883}
\newcommand{\amplitudetotaljointslopemodel}{0.116}

\newcommand{\rmsjointslopemodel}{0.343}
\newcommand{\stdmeandailyrvchromaticcorrectedred}{2.03}

\newcommand{\rmsrvsdcpca}{1.30}
\newcommand{\rmsrvsfiesta}{1.03}
\newcommand{\rmsrvsgausshermite}{1.29}
\newcommand{\rmsrvsscalpelsfive}{0.981}
\newcommand{\rmsrvsscalpels}{0.277}
\newcommand{\rmsrvsCaHKII}{1.39}
\newcommand{\rmsrvsHalpha}{1.48}
\newcommand{\rmsrvsbis}{1.66}
\newcommand{\rmsccfarea}{0.00056}
\newcommand{\cvrmsrvsdcpcafivepre}{1.34}
\newcommand{\cvrmsrvsdcpcafivepost}{1.32}
\newcommand{\cvrmsrvsdcpcaninepre}{1.31}
\newcommand{\cvrmsrvsdcpcaninepost}{1.21}

\newcommand{\cvrmsrvsfiestafivepre}{1.21}
\newcommand{\cvrmsrvsfiestafivepost}{1.06}

\newcommand{\cvrmsrvsgausshermitetenpre}{1.15}
\newcommand{\cvrmsrvsgausshermitetenpost}{1.11}
\newcommand{\cvrmsrvsscalpelsfivepre}{0.969}
\newcommand{\cvrmsrvsscalpelsfivepost}{0.784}

\newcommand{\cvrmsrvsscalpelstenpre}{0.510}
\newcommand{\cvrmsrvsscalpelstenpost}{0.426}
\newcommand{\cvrmsrvsscalpelstenoneoffour}{0.350}
\newcommand{\cvrmsrvsscalpelstentwooffour}{0.422}
\newcommand{\cvrmsrvsscalpelstenthreeoffour}{0.340}
\newcommand{\cvrmsrvsscalpelstenfouroffour}{0.373}

\usepackage[T1]{fontenc} 
\usepackage[utf8]{inputenc} 

\newcommand{\PSUAA}{Department of Astronomy and Astrophysics, Penn State University, 525 Davey Laboratory, 251 Pollock Road, University Park, PA, 16802, USA}
\newcommand{\PSUCEHW}{Center for Exoplanets and Habitable Worlds, Penn State University, 525 Davey Laboratory, 251 Pollock Road, University Park, PA, 16802, USA}
\newcommand{\PSETI}{Penn State Extraterrestrial Intelligence Center, Penn State University, 525 Davey Laboratory, 251 Pollock Road, University Park, PA, 16802, USA}
\newcommand{\UA}{Steward Observatory, University of Arizona, 933 N.\ Cherry Ave, Tucson, AZ 85721, USA}

\newcommand{\Penn}{Department of Physics and Astronomy, University of Pennsylvania, 209 S 33rd St, Philadelphia, PA 19104, USA}

\newcommand{\noirlab}{U.S. National Science Foundation National Optical-Infrared Astronomy Research Laboratory, 950 N.\ Cherry Ave., Tucson, AZ 85719, USA}

\newcommand{\JPL}{Jet Propulsion Laboratory, California Institute of Technology, 4800 Oak Grove Drive, Pasadena, California 91109}

\newcommand{\UCI}{Department of Physics \& Astronomy, The University of California, Irvine, Irvine, CA 92697, USA}
\newcommand{\Carleton}{Carleton College, One North College St.,  , MN 55057, USA}
\newcommand{\Carnegie}{Earth and Planets Laboratory, Carnegie Science, 5241 Broad Branch Road, NW, Washington, DC 20015, USA}
\newcommand{\PSUICS}{Institute for Computational and Data Sciences, Penn State, University Park, PA, 16802, USA}
\newcommand{\PSUCASt}{Center for Astrostatistics, Penn State University, 525 Davey Laboratory, 251 Pollock Road, University Park, PA, 16802, USA}

\newcommand{\DTU}{DTU Space, National Space Institute, Technical University of Denmark, Elektrovej 328, DK-2800 Kgs. Lyngby, Denmark}
\begin{document}


\title{Earths within Reach:  Evaluation of Strategies for Mitigating Solar Variability using 3.5 years of NEID Sun-as-a-Star Observations}


\author[0000-0001-6545-639X]{Eric B.\ Ford}
\affil{\PSUAA}
\affil{\PSUCEHW}
\affil{\PSUICS}
\affil{\PSUCASt}


\author[0000-0003-4384-7220]{Chad F.\ Bender}
\affil{\UA}

\author[0000-0002-6096-1749]{Cullen H.\ Blake}
\affil{\Penn}

\author[0000-0002-5463-9980]{Arvind F.\ Gupta}
\affil{\noirlab}

\author[0000-0001-8401-4300]{Shubham Kanodia}
\affil{\Carnegie}

\author[0000-0002-9082-6337]{Andrea S.J.\ Lin}
\affil{\PSUAA}
\affil{\PSUCEHW}

\author[0000-0002-9632-9382]{Sarah E.\ Logsdon}
\affil{\noirlab}

\author[0000-0002-4927-9925]{Jacob K. Luhn}
\affil{\UCI}

\author[0000-0001-9596-7983]{Suvrath Mahadevan}
\altaffiliation{Principal Investigator}
\affil{\PSUAA}
\affil{\PSUCEHW}

\author[0000-0002-4677-8796]{Michael L. Palumbo III}
\affil{\PSUAA}
\affil{\PSUCEHW}

\author[0000-0002-4788-8858]{Ryan C. Terrien}
\affil{\Carleton}

\author[0000-0001-6160-5888]{Jason T.\ Wright}
\affil{\PSUAA}
\affil{\PSUCEHW}
\affil{\PSETI}

\author[0000-0001-5290-2952]{Jinglin Zhao}
\affil{\DTU}
























\author[0000-0003-1312-9391]{Samuel Halverson}
\affil{\JPL}

\author[0009-0007-3512-7705]{Emily Hunting}
\affil{\noirlab}

\author[0000-0003-0149-9678]{Paul Robertson}
\altaffiliation{Instrument Team Project Scientist}
\affil{\UCI}

\author[0000-0001-8127-5775]{Arpita Roy}
\affiliation{Astrophysics \& Space Institute, Schmidt Sciences, New York, NY 10011, USA}


\author[0000-0001-7409-5688]{Guðmundur Stefánsson} 
\affil{Anton Pannekoek Institute for Astronomy, University of Amsterdam, Science Park 904, 1098 XH Amsterdam, The Netherlands}



\begin{abstract}
We present the results of Sun-as-a-star observations by the NEID Solar Telescope at WIYN Observatory, spanning January 1, 2021 through June 30, 2024.  
We identify 117,060 observations 
which are unlikely to be significantly affected by weather, hardware or major calibration issues.  
We describe several high-level data products being made available to the community to aid in the interpretation and inter comparisons of NEID solar observations.  
%
Solar observations demonstrate excellent performance of NEID, including radial velocity (RV) accuracy and long-term stability of better than $\simeq~0.37$\mps\ over $\simeq$~3.5 years, even though NEID was not originally designed or optimized for daytime observations of the Sun. 
%
Currently, intrinsic stellar variability is the primary barrier to detecting Earth-analog planets for most nearby, Sun-like stars.  
We present a comparison of the effectiveness of several methods 
proposed to mitigate the effects of solar variability on the Sun's estimated radial RV. 
We find that the Scalpels algorithm performs particularly well and substantially reduces the root mean square (RMS) solar RV from over 2\mps\ to \rmsrvsscalpels~ \mps.
Even when training on a subset of days with NEID solar observations and testing on a held-out sample, the RMS of cleaned RV is \cvrmsrvsscalpelstenthreeoffour--\cvrmsrvsscalpelstentwooffour~ \mps.
This is significantly better than previous attempts at removing solar variability and suggests that the current generation of EPRV instruments are technically capable of detecting Earth-mass planets orbiting a solar twin if provided with sufficient observing time allocations ($\sim~10^3$ nights of observations).
\end{abstract}
  

\section{Introduction}
\label{sec:intro}
\subsection{NEID Solar Observations}
\label{sec:intro_neid}
NEID is a highly-stabilized, high-resolution spectrograph designed to conduct extremely precise radial velocity (EPRV) measurements during night-time observations from the WIYN 3.5m observatory\footnote{The WIYN Observatory is a joint facility of the NSF's National Optical-Infrared Astronomy Research Laboratory, Indiana University, the University of Wisconsin-Madison, Pennsylvania State University, Purdue University and Princeton University.} \citep{NEID_optical}.  
An auxiliary NEID Solar Telescope allows for routine daytime observations of the Sun \citep{NEID_solar_hardware}.
While the spectrograph and detector are shared between night-time and solar observations, there are some important differences in the light path and instrument stability between daytime solar and night-time stellar observations (e.g., longer fiber from the solar telescope, finite angular size of the sun, different optical path and calibration, no atmospheric dispersion corrector, pointing errors are driven by the solar tracker and not the WIYN 3.5m telescope) that we will discuss in \S\ref{sec:results}.  

Most aspects of NEID, the official data reduction pipeline (DRP)\footnote{\url{https://neid.ipac.caltech.edu/docs/NEID-DRP/}}, 
and the detailed error budget \citep{NEID_budget} apply to both night-time and solar observations\footnote{The NEID instrumental error budget is 0.27\mps \citep{NEID_budget}.  However, there was not a formal error budget or requirement for the NEID Solar Telescope, since it was an auxiliary subsystem. 
Since a large fraction of the light path and hardware is shared, we do not expect that the error budget of solar observations will vastly exceed that of nighttime observations.}.
The typical formal measurement uncertainty reported by the NEID DRP for each NEID solar spectrum is 0.285 \mps, not significantly different than for a bright nighttime star.  
  
However, unlike nighttime star observations, it is practical to obtain hundreds of solar observations every clear day,  providing the opportunity to bin observations to obtain very high signal-to-noise, $\sim2,000$ which may be useful for mitigating the impact of stellar line shape changes \citep{DavisPca,PalumboGrassII}. 
If there were no contamination from stellar variability or other sources of systematic ``noise'' (e.g., telluric variability, instrumental variability), then the information content of one hour of NEID solar observations could achieve a precision of $\sim0.05$\mps and daily averaged NEID solar RVs could achieve a precision of $\sim$0.02\mps.

In practice, the root mean square deviation (RMS) of daily-averaged RVs observed by NEID (and other solar telescopes) is multiple \mps.
The specific value depends on the dates where observations were taken, as well as the the quality of the instrument calibration and data reduction methods.  
The large number of solar observations (and favorable observing cadence) could prove valuable for evaluating strategies for mitigating the impact of stellar variability on EPRV surveys (see \S\ref{sec:intro}) and for empirically constraining some terms of the instrumental error budget and evaluating the effectiveness of various methods proposed for mitigating them.

Several planet-hunting spectrographs are collecting solar observations \citep{HARPSNSolarTelescope,NEID_solar_hardware,EXPRESSolarTelescope,MaroonX,KPFSolarTelescope,LOCNESSolarTelescope}.  
NEID solar observations offer a unique combination of features, including high spectral resolution (R$\sim$120,000), an ultra-stabilized spectrograph enclosure \citep{Robertson_environment,NeidEnvControl}, an unusually large wavelength grasp ($\le$380--$\ge$930nm), rapid observing cadence (55s exposures with gaps for readout lasting 28s (38s) starting (before) September 17, 2021), a dataset that includes both low and high levels of solar magnetic activity, and data that become public promptly after processing.  
Thus, we envision the NEID solar observations as providing a valuable community resource for quantifying the impact of intrinsic stellar variability on the sensitivity of future EPRV surveys to detect and/or characterize low-mass planets.  
This paper presents NEID solar observations collected during January 2021---June 2024 and provides high-level data products that will be useful for future analyses.

\subsection{Evaluating Stellar Variability Mitigation Strategies}
\label{sec:intro_eval_mitigation}
Thanks to the dramatic improvements in spectrographs and calibration systems, stellar variability is often the dominant source of ``noise'' limiting the effective detection sensitivity of EPRV surveys \citep{EPRVWGReport}.  
Most methods for measuring ``the radial velocity of a star'' assume that the true stellar spectrum remains constant in the star's center-of-mass frame \citep{HaraFordReview}.  
From this perspective, line-shape changes in the stellar spectrum can be seen as an additional source of astrophysical ``noise'' that contaminates measured RVs.  
Several previous studies have attempted to clean the observed RVs of this additional noise, based on other changes in individual spectra (e.g.,  classical spectroscopic activity indicators, data-driven variability indicators) and/or the temporal structure in measured RVs.
While several approaches appear to help, the community has not yet identified a single strategy that can be counted on to enable the detection of an Earth  mass-planet orbiting in the Habitable Zone of a Sun-like star.

One challenge in evaluating stellar variability mitigation strategies is the results from each method inevitably depend on many additional details of the data and analysis, such as which instrument was used, the pipeline (and version) used for extracting spectra from raw observations, the dates of the observations (due to constantly changing extent and distribution of magnetically active regions), how the researchers decide which observations should be included/excluded in the analysis, and the choice of which stellar features to include in the analysis.
Therefore, it is often not straightforward to compare the effectiveness of two proposed strategies based on the results published in two separate studies.  

To address this and related challenges, the EPRV community has organized multiple data challenges to help researchers compare different methods to search for planetary signals \citep{DumusqueChallengeI,DumusqueChallengeII}, to estimate the evidence for planetary detections \citep{NelsonEvidenceChallenge}, and to reduce the impact of stellar variability on RV measurements \citep{ZhaoEsspData,ZhaoEsspTwoComparison,ZhaoEsspThree}.   
\citet{DumusqueChallengeI} provided data challenge participants with just a few potential indicators, while \citet{ZhaoEsspData} provided extracted spectra and let participants apply a wider variety of strategies to mitigate stellar variability.
For the latter study, eleven research teams submitted their attempts to clean the measured RVs of stellar variability using 22 methods.  
While many of the methods succeeded in reducing the RMS of cleaned RVs, the different methods disagreed on the magnitude (and sometimes the sign) of the correction that was needed at each time \citep{ZhaoEsspTwoComparison}.
The differences were significant and demonstrated that further research is needed to determine which methods (if any) are powerful, robust, and reliable for detecting low-mass planets.  
While the EXPRES Stellar Signals Project (ESSP) Data Challenge was valuable, the interpretation of the results was complicated by the use of real observations of stars for which the true number, masses, and orbits of planets are unknown \citep{ZhaoEsspTwoComparison}.  
Not having access to the ground truth limited how precisely the data challenge could evaluate and compare the various mitigation strategies.

Solar observations provide a unique opportunity to evaluate strategies for mitigating RV ``noise'' from stellar variability, since the relative motion between the Sun and observatory is known precisely.  
In this study, we apply multiple strategies for mitigating stellar variability and compare their effectiveness using a common NEID solar dataset.  
%

%
This paper presents results of analyzing solar observations and describes various data release products to ease the analysis of NEID solar observations for other researchers.  
It also presents a comparison of several wavelength-domain methods for characterizing the instantaneous level of stellar variability and its impact on RV measurements.  
The manuscript is organized as follows.  
In \S\ref{sec:filter}, we describe a series filters applied to select high-quality observations.  
In \S\ref{sec:methods}, we describe our methods for analyzing extracted spectra to calculate cross correlation functions (CCFs), radial velocities (RVs), and CCF bisectors.
In \S\ref{sec:results}, we  describe the high-level data products that we provide and present results from their first analysis.
In \S\ref{sec:eval_mitigation}, we evaluate and contrast the ability of several wavelength-domain strategies for mitigating solar variability.  
In \S\ref{sec:discuss}, we discussion the implications of our findings and opportunities for future improvements. 


\section{Filtering of Observations}
\label{sec:filter}
Before performing a scientific analysis of NEID solar observations, one needs to filter out observations which were corrupted by weather, known hardware issues, and significant wavelength calibration issues.  
We describe a series of filters that are useful in constructing a clean dataset for subsequent analysis.  
We start with essential filters and work towards less common, smaller, and/or less well understood effects.

\subsection{Essential filters}
Here, we describe several filters that we consider essential and recommend all subsequent analyses of NEID solar observations perform.

\subsubsection{Wavelength calibration applied}
\label{sec:filter_wavecal}
Since EPRV measurements are differential, it is important to only compare RVs based on data reduced using the same major and minor versions of the NEID DRP.  This paper presents results based on NEID DRP v1.3.  
Since solar observations were paused at the end of June 2024, no solar observations were affected by issues necessitating the bug fixes included in DRP v1.3.1.

\begin{figure}
    \centering
    \includegraphics[width=0.5\linewidth]{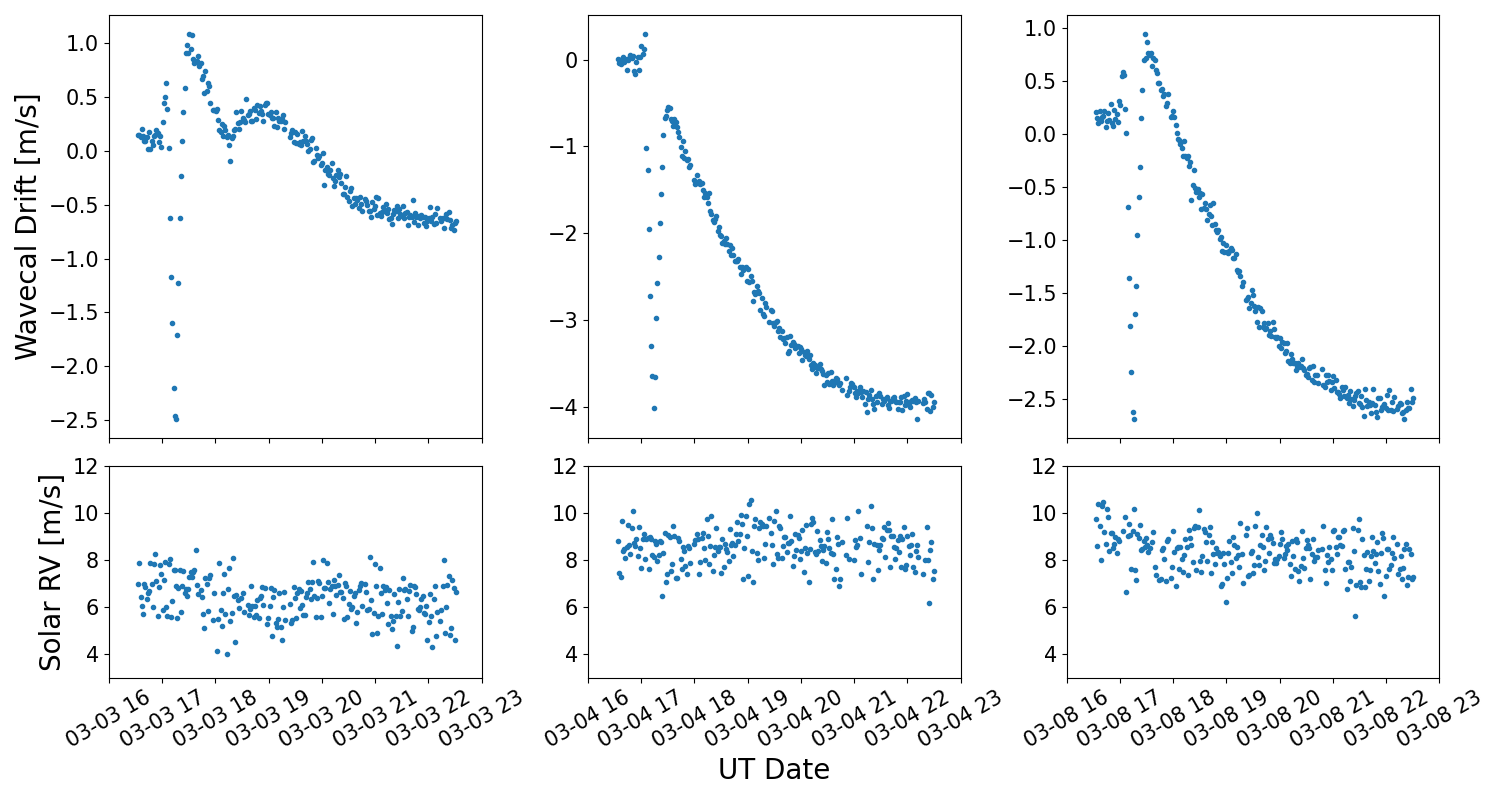}
    \caption{Top:  Measured instrumental drift over the course of solar observations for each of three days.  Bottom:  Measured solar RVs after accounting for measured instrumental drift.  The NEID DRP does an excellent job of removing the short-term transients in wavelength calibration following the daily refill of the liquid nitrogen dewar.  The  RMS RV from a smoothed version of the instrument drift (after the initial large transient) is 0.052--0.055 \mps.  
    However, an observed trend in solar RVs over the course of each day is described in \S\ref{sec:daily_slope}. 
    The solar RVs are potentially subject to additional systematic effects not probed by the etalon in the calibration fiber. These effects could include pointing, coupling-related behaviors with the solar feed or sci-cal fiber drift, and differential extinction. \label{fig:ex_daily_drift} }
\end{figure}

Since we aim to focus our attention on high-quality observations, we filter for observations with a wavelength calibration based on a combination of the laser frequency comb (LFC) in the red and Thorium-Argon lamp in the blue and that use the \verb|dailymodel0| drift function.  
These are indicated by the FITS header keywords WAVECAL and DRIFTFUN of the FITS files provided by the NEID DRP (level 0, 1 and 2).
\verb|dailymodel0| is the default drift model in DRP v1.1-v1.3 which compares Fabry-Perot (FP) etalon measurements which are taken simultaneously with the solar exposures to the daily wavelength solution derived from the standard NEID dawn/dusk calibration sequences.
\footnote{See the NEID Data Reduction Pipeline documentation at \url{https://neid.ipac.caltech.edu/docs/NEID-DRP/algorithms.html\#wavelength-calibration-sources} for details.}
Despite efforts to stabilize NEID's environment, the instrument does experience a significant RV drift during the daylight hours due to the daily filling of the liquid nitrogen dewar (see Figure \ref{fig:ex_daily_drift}).
Most of the drift is removed as part of NEID's DRP using the \verb|dailymodel0| drift function that combines information from calibration observation taken at the beginning (end) and end (beginning) of each night (day) of observing.  
The NEID wavelength calibration procedure and wavelength calibration model was designed to maximize accuracy of night-time observations on stars.
Unfortunately, that means that there are large, non-linear changes in the wavelength calibration during solar observations.  
In \S\ref{sec:daily_slope} we show that there is room for further improvements in the daily drift model for solar observations.  

\subsubsection{Hardware issues}
On three occasions, and likely due to inclement weather, the NEID solar telescope cable bundle became dislodged from its optimal wrap path. During these windows, NEID solar observations were adversely affected as a result of tension on the cable bundle causing very poor tracking.  
The NEID fiber scrambling system minimizes the effects of pointing errors \citep{kanodia2023} and an oversized input port in the integrating sphere further suppresses the impact of pointing errors on NEID solar observations \citep{NEID_solar_hardware}.  
However, the pointing errors due to the cable bundle issue were so extreme that they still resulted in a substantial RV signal, likely due to a truncation of the stellar disk.   
Therefore, we exclude data taken during the intervals October 2-27, 2021, 
February 5-27, 2023, and March 18-31, 2023.  
Initially, these were attributed to bad weather events, but after multiple occurrences, the root cause of the problem was found and the cable bundle was more securely attached to the roof in April 2023. We do not anticipate additional data gaps due to this issue.

\subsubsection{Contreras fire}
The Contreras Fire led to a pause in NEID solar operations, with the last day of pre-fire solar observations being June 13, 2022.  
The spectrograph was warmed up in order to mitigate severe risks to the hardware stemming from uncontrolled thermal shocks.
Once the fire was put out and facilities recovered, NEID resumed both solar and nighttime operations in low precision mode on November 12, 2022, while the instrument was still cooling down and settling to its new state.  Some high-precision night time observations resumed on November 23, 2022, and the NEID Guaranteed Time Observations program resumed December 1, 2022.  
While one could likely recover additional observations during the cool down period with custom attention, we will exclude data taken after the fire and prior to December 1, 2023.
After applying these filters (plus excluding solar eclipse days \S\ref{sec:filter_solar_eclipse}, we are left with 192,373 ``available'' solar observations, $\simeq84$\% of all NEID solar data collected.

We define data taken during January 1, 2021 -- June 13, 2022 as ``run 1'', and during December 1, 2022 -- June 30, 2024 as ``run 2''.
Starting July 1, 2024 solar observations were paused while the etalon simultaneous calibration source was unavailable for monitoring instrumental drift.  

\subsubsection{Solar telescope on target}
\label{sec:filter_onsky}
Solar observations are scripted and automated to run from 16:30 to 22:30 UT.  
We found that the last solar observations of a day were sometimes impacted (e.g., solar telescope starts slewing to stow position before final exposure completes, potentially a partial obscuration or specular reflection).  
Therefore, we restrict our analysis to observations which start between 16:30:00 and 22:12:00 UT each day.  It is likely that about a dozen additional clean observations could be recovered on many clear days, but additional effort would be required.
This filter resulted in losing 5.5\% of available observations.

\subsubsection{Weather}
\label{sec:filter_weather}
Next, we apply a series of filters designed to exclude observations potentially contaminated by poor observing conditions.  
The NEID solar telescope takes observations even when the Sun is partially obscured by clouds.  
Since the equatorial limbs of the sun are rotating at nearly 2 k\mps, even small variations in transmittance as a function of position on the disk can cause a substantial spurious RV.  
NEID has a dedicated exposure meter that picks off light coming through the solar fiber and reads out once a second.
The NEID solar tracker also has an independent pyrheliometer that records the instantaneous total solar irradiance during observations in the level 0 files from the NEID DRP.  The pyrheliometer has a substantially larger field of view ($\sim$5 degrees) than the telescope assembly and is more sensitive to infrared radiation that is susceptible to absorption by water vapor. 

We require a minimum flux (averaged over the exposure) from both the pyrheliometer (pyrflux\_mean $\ge 10^{2.95}$ W/m$^2$) and exposure meter (expmeter\_mean $\ge 10^5$ ADU).   
Fluctuations in either the exposure meter or pyrheliometer reading during an exposure suggest transparency variations in the vicinity of the Sun could affect observations.
Therefore, we compare the mean (expmeter\_mean) and root mean square (RMS) deviation from the mean of the exposure meter during each exposure (expmeter\_rms) and require that 
expmeter\_rms $\le 0.003 \times$expmeter\_mean.  
Similarly, we compute the mean 
and RMS deviation from the mean duration each exposure (pyrflux\_rms) and require 
pyrflux\_rms $\le 0.0035 \times$ pyrflux\_mean.
While the precise values could be tweaked, we have found these thresholds to be useful for a preliminary NEID data release based on January 2021--June 2022\footnote{\url{10.5281/zenodo.7857521}} and found no reason to change them.
The weather filters resulted in losing an additional 33\% of available observations.

\subsection{Additional filters}
Here, we describe several filters that we found were valuable for the current analysis.  
However, we recognize that other researchers may want to adapt different threshold values depending on their specific science goals.

\subsubsection{Pointing/tracking/fiber coupling}
\label{sec:filter_pointing}
The minimum exposure meter flux provides some check on the pointing.  
However, we found a more powerful filter to be based on the ratio of the flux from the exposure meter and the pyrheliometer.  
We note that this is the only filter which we have updated since the original recommendations developed based on run 1.  
Previously, we filtered for expmeter\_mean / ADU $\ge 150 \times\,$pyrflux\_mean /(W/m$^2$) in order to minimize risk of any undetected pointing errors.
However, the system throughput decreased between run 1 and run 2 by nearly 14\% (as measured by the exposure meter).  Therefore, we have replaced this filter with one that requires that the ratio  expmeter\_mean/pyrflux\_mean exceed 95\% of the median value for that ratio within each day. This filter resulted in losing less than 0.2\% of available observations. 

\subsubsection{Airmass}
\label{sec:filter_airmass}
Differential atmospheric extinction can induce a spurious RV signal.  
For this analyses, we restrict our attention to observations with an airmass less than 2.5.
This filter resulted in losing less than 1\% of available observations.
For some science cases, it may be useful to limit observations to those taken at lower airmass, so as to limit the impact of by differential extinction.  

\subsubsection{Binning \& Minimum number of observations}
\label{sec:filter_binning}
In order to reduce the RV impact of solar oscillations, we will focus on analyzing binned observations.
For the analyses presented below, we will work primarily with binned observations, requiring five consecutive exposures (within the same day) to pass the above criteria.  
This corresponds to $5\times~55$s integration time and a wall clock time of 7.12 minutes (for early observations) or 6.45 minutes (starting September 16, 2021).   
Working from binned observations has the additional benefit of reducing the risk of weather impacting RVs, since points passing thresholds by chance are unlikely to be consecutive.  
Our analysis of interday RV and CCF variations will focus on days with at least 15 complete binned observations to further reduce the impact of oscillations, reduce the chance of observations during poor atmospheric conditions, and ensure a high signal-to-noise ratio.

Depending on one's science goals, other analyses might choose to bin data differently (e.g., characterizing pulsations), so we provide a table with each unbinned observation as well as a table binning 10 consecutive exposures.
Since we have not inspected unbinned observations for data quality issues, many users will likely prefer to work from the binned observations. 
Similarly, we provide data for all days with at least one binned observations, but have not performed a detailed inspection of data quality for days with less than 15 binned observations.

\subsubsection{Wavelength calibration issues}
\label{sec:wavecal_filter}
In \S\ref{sec:methods} we will describe how we calculate RVs from each order of the spectrum (with sufficient lines, signal-to-noise, and accurate wavelength calibration).  
By comparing the order RVs within each observation, we can recognize days where there is an unusually large dispersion of derived RVs across orders, most likely due to calibration issues (e.g., LFC flux, pipeline configuration).  
Most of these outliers can be tracked to a small number of orders near the transition between orders with and without LFC calibration.
This has been traced to sporadic episodes of low flux from the LFC in the bluest orders where the NEID DRP uses the LFC for wavelength calibration.  
While this primarily affects echelle orders 117 \& 118, thanks to the large number of solar observations, we can also recognize it sometimes affecting echelle orders 113-118.  
While we provide tables with the RVs extracted from these orders, we exclude these orders for computing CCFs and RVs for subsequent analysis in this paper.  
With adequate investment, a future version of the NEID DRP could likely recover additional RV information from these orders.

\subsubsection{Solar eclipses}
\label{sec:filter_solar_eclipse}
NEID observed partial solar eclipses on October 14, 2023 and April 8, 2024.  
These provide a unique opportunity for obtaining spatially resolved information with the NEID spectrograph and will be the subject of a future publication (Gonzales et al., in prep).  
We exclude these days from plots and summary statistics.

\subsubsection{Outcome of filtering}
\label{sec:filter_overview}
We computed CCFs and RVs for 129,889 NEID solar observations. 
After applying the above filters, we are left with 
\numobsused~ NEID solar observations and \numbinnedobsused~ binned observations that include all five observations.  
There are \numdaysmeanrv~ days with at least one complete binned observation and \numdaysmeanslope~ days with at least 15 binned RVs to be used for subsequent analysis.


\section{Analysis Methods}
\label{sec:methods}

\subsection{Penn State Research Pipeline}
\label{sec:psrp}
In order to help evaluate various strategies for analyzing EPRV observations, we have developed the Penn State Research Pipeline (PSRP).  
The PSRP complements and does not replace existing instrument-specific data reduction pipelines, as it starts from order-extracted spectra.
This way, PSRP makes use of instrument-specific knowledge that instrument teams have incorporated into their data reduction pipelines.
PSRP adds value by providing multiple alternatives for computing CCFs, estimating RVs, and applying data-driven methods for mitigating stellar variability.  
For the official NEID DRP, reliability and stability of the pipeline are essential.  
In contrast, PSRP prioritizes flexibility, computational speed, and a modular and composable design, so as to allow researchers to rapidly try out multiple algorithms and parameter values.

The PSRP is based on a combination of Julia packages that are part of the \verb|RvSpectML| organization.
\verb|RvSpectML| provides many options for processing of spectral time series.  
Here we provide an overview of the methods used for the analysis presented in this paper.

\subsubsection{Ingesting FITS files}
\label{sec:methods_io}
\texttt{EchelleInstruments.jl} provides a common interface for reading individual order extracted spectra and providing information specific to each instrument.
%
We perform a blaze normalization based on the blaze function estimated for each observation by the NEID DRP.
Next, we perform continuum normalization that closely follows the Rassine algorithm \citep{Cretignier_Rassine}.
The anchor points are computed from the summed spectrum over one clear day and then reused for continuum normalizing all other days.

\subsubsection{CCFs}
\label{sec:methods_ccfs}
\texttt{EchelleCCFs.jl} computes the cross-correlation function between each observed spectrum and a mask.  
Each mask is specified by list of entries (each including the order, lines, and weight) and the mask shape.  
We present analyses based on two line lists.  
One is based on the ESPRESSO G2 mask, so as to facilitate comparisons to previous work.  
Since the ESPRESSO mask is limited to 380-786nm, we also compute a custom line list to use as much of NEID's wavelength grasp as practical based on \texttt{RvLineList.jl}.  
In both cases, stellar lines can be rejected due to being near telluric absorption lines based on the predicted slope in atmospheric transmittance (as a function of velocity) as  computed by TAPAS \citep{tapas}.  

For both line lists, our baseline analysis presented here uses a top-hat mask with width of 620.953 \mps and the CCF was evaluated at velocities within 12.5 \kmps of a reference velocity roughly centered on the solar RV.  
In addition to the CCFs for each observation, we also compute daily averaged CCFs that are particularly useful for studying rotationally-linked stellar variability.  
While we tabulate daily averaged CCFs for days with only a few observations, we recommend focusing on days with at least 15 binned observations and follow this recommendation in our analysis below.  

\subsection{Radial Velocities}
\label{sec:methods_rvs}
\verb|RvSpectML.jl| provides multiple algorithms for estimating RVs from CCFs.
For this study, RVs are measured separately for each order by fitting a Gaussian line profile to each order CCF.  
For each order, we compute RVs from the individual observations as well as daily averaged RVs.   
For intraday RV analyses (\S\ref{sec:rms_rv_within_day} and earlier), order weights are proportional to the inverse mean (over days with at least 15 binned observations passing the filter above) of the mean squared deviation of the binned order RVs from that day's mean RV.  
For comparing RVs across days, we compute an RV offset for each order and each observing run based on the daily means.
The run offsets are subtracted from each order RV and then we add back the weighted mean RV offset for each observation's run based on the mean of a set of reliably LFC-calibrated orders (physical orders 98-112).  
Once RV offsets have been computed for each (run, order) pair, a single RV is estimated for each observation by taking a weighted average of order RVs.
At this stage, updated order weights are computed proportional to the inverse of the mean squared deviation of the daily binned order RVs from their mean (over days with at least 15 binned observations).  
Orders with known wavelength calibration issues (see \S\ref{sec:wavecal_filter}) and large measurement uncertainties are assigned zero weight.

\subsection{CCF bisectors}
\label{sec:methods_biss}
We compute a weighted sum (over orders) of daily binned order CCFs, using weights proportional to the inverse of the mean squared deviation of the daily binned order RVs from their mean (over days).  
Orders with known wavelength calibration issues (see \S\ref{sec:wavecal_filter}) are assigned zero weight.
To compute the velocity at each of several target CCF depths (on both sides of the minimum), we interpolate using a Matern-$\frac{5}{2}$ Gaussian process conditioned on the observed CCF.  
We average and subtract the values from the left and right side of the CCF to obtain the CCF bisector and CCF velocity span at each target depth.
We estimate RVs separately from the top and bottom portion of the daily average CCF. 
We compute a bisector inverse slope (BIS), here defined as the difference of the RV measured from the top of the CCF ($RV_{\mathrm{top}}$, using fractional depths of 10--40\%) and the RV measured from the bottom of the CCF ($RV_{\mathrm{bottom}}$ using fractional depths of 55--95\%).
We compute the bisector curvature as $(v_{top}-v_{mid})-(v_{mid}-v_{bottom})$.
For estimating the curvature, the top is defined as the mean of the bisectors between depths of 30 and 50\%, the middle is based on depths from 50 and 75\%, and the bottom is based on depths from 75 and 90\%.  
In each case, bisector is measured relative to the best-fit CCF normalization at depths spaced by 0.01.  
The depths used for measuring the BIS and CCF curvature have been adjusted slightly from their original definitions in the literature based on the observed shape of CCF and robustness of measurements at shallow CCF depths.  

\subsection{Stellar activity indicators}
The NEID DRP provides measurements of several classical activity indicators (e.g., Ca II H\&K and H$\alpha$) in FITS headers.  
Our tables merely propagate these values to enable their convenient analysis.
Figure \ref{fig:activity_vs_time} shows time series for daily averages of four indicators that appear likely to be the most useful for characterizing rotationally-modulated variations and long-term activity variations.  

\subsection{Additional solar-specific corrections}
\label{sec:solar_corrections}
{\em Data quality summary statistics:}  
We compute several additional quantities that can be useful when analyzing Sun-as-a-star observations.  
Summary statistics based on the exposure meter and pyrheliometer were described in \S\ref{sec:filter_weather} \& \ref{sec:filter_pointing}.
The NEID DRP already provides the Doppler boost necessary to transform observations into a (fictitious) frame such that the true center-of-mass solar velocity would be a constant in FITS headers SSBZNNN, where NNN specifies the echelle order \citep{WrightBarycorpySolar}.
We compute differential extinction and 
apparent rotation velocity using \verb|NeidSolarScripts.jl|.

{\em Differential extinction:}  
Next, we estimate the apparent Doppler shift due to differential extinction, $\Delta$v\_diff\_ext or $\Delta v_r$, following notation of \citet{ColierCameron2019}. 
$\Delta v_r$ can be expressed as the product of the extinction coefficient, $k$, and a complex function of geometry and a limb darkening coefficient (set to 0.6).  
Since NEID is able to identify measurements affected by transparency variations using data from the exposure meter and pyrheliometer individually (as opposed to in a statistical sense), we do not need to employ a mixture model like that of \citet{ColierCameron2019} for estimating $k$.  Instead, we simply fit a linear model to the signal-to-noise in echelle order 98 ($619.5--631.4$ nm) \citep[which has similar wavelengths to HARPS-N ``order 60'' used by][]{ColierCameron2019} versus airmass.
Modeling differential extinction as a single, grey shift is an approximation to be improved in a future version of the PSRP.  
Improving this approximation could be particularly important for making use of NEID's far red and NIR orders.

One complication is that there are significant changes in the signal-to-noise due to imperfect coupling of the solar fiber to the science fiber.  
The solar observations are scripted to periodically delay the start of exposure, so the port fiber can realign and recenter the solar fiber on the NEID science fiber, leading to occasional jumps in the SNR \citep{NeidPortAdapter}.
We use the timestamps of all NEID solar observations within a day (even those unusable due to weather or pointing) to identify when the port fiber aligner makes adjustments.
We group observations taken with gaps between consecutive exposure start times of less than 100s as belonging to a single ``segment''.
When fitting for each day's parameters, we allow each segment to have its own SNR offset, but require all segments to share a common extinction coefficient.

There can also be changes in the throughout near solar noon, when the solar telescope flips to position itself for afternoon observations.    
Therefore, we exclude observations starting within 15 minutes of solar noon when fitting a differential extinction coefficient.  

While a simple linear fit works well on days with plenty of data taken in clear weather, it is not practical to make a good measurement for some days with few clear observations. 
Therefore, we apply Bayesian inference, adopting priors that guide the model parameters to reasonable values on days with few observations.  
We adopt diffuse Normal priors for each segment intercept with zero mean and unit variance.
We adopt a log Normal prior for the extinction coefficient with $\mu=\ln(0.1)+0.5^2$ and $\sigma=0.5$.
This results in $k\simeq 0.1 $ on days with few usable observations, likely leading to underestimating the magnitude of $\Delta v_r$ on such days.
For the jitter parameter, $\sigma_j$ we adopt a log Normal prior distribution with $\mu=\ln(0.05)+2^2$ and $\sigma=2$.
We use Markov chain Monte Carlo (MCMC) with the No U-Turn Sampler (NUTS) to obtain a posterior sample, the  mean, median, and standard deviation of the marginal posterior for $k$, and credible intervals.

Given the potential for future work to provide updated differential extinction coefficients, we tabulate corrections $\Delta v_r$ based on formalism of \citet{ColierCameron2019} and a constant extinction coefficient of $k_\circ=0.157482$ (the value of a preliminary fit derived from a subset of the NEID solar dataset, done as part of a preliminary analysis).  
When applying differential extinction corrections for a given day, we first rescale the tabulated $\Delta v_r$ by the ratio $k_{\mathrm{med}}/k_\circ$, where $k_{\mathrm{med}}$ is the median of the marginal posterior for $k$ for that day based on MCMC.   

{\em Apparent rotation velocity:}  
The apparent solar angular rotation rate varies due to Earth's rotation and eccentric orbit.  
These affect the measured line widths and propagates to affect the CCFs.  
We calculate the difference between the squares of the full width half maximum (FWHM) of the CCF as observed and the FWHM of the CCF if it were observed in a sidereal frame ($\Delta 
\mathrm{fwhm}^2 = F_{\mathrm{obs}}^2 - F_{\mathrm{sid}}^2$) following conventions of \citet{ColierCameron2019}
We adopt the same scaling factor $\gamma=1.04$ estimated by \citet[][]{ColierCameron2019}.  
Note that this field is in km  s$^{-1}$ instead of \mps.  
Since flux is preserved, the depths of lines and CCF contrast are also affected.  We calculate a sidereal CCF contrast, 
$C_{\mathrm{sid}} = C_{\mathrm{obs}} \frac{F_{\mathrm{obs}}}{F_{\mathrm{sid}}}$.  

For each daily averaged CCF, we fit a Gaussian CCF profile  allowing for an unknown normalization, location, width, and depth.   
The observed FWMH is estimated based on the best-fit parameters assuming a Gaussian CCF shape.   
The observed contrast is estimated as the product of the best-fit normalization and depth.
Results are shown in the left panel of Figure \ref{fig:ccf_metrics_vs_time}.  
We find that the product CCF area, $A = C_{\mathrm sid} \times F_{\mathrm{obs}}$ is nearly constant, so $A$ and $C_{\mathrm{sid}}$ are not shown.  
The RMS of daily-averaged CCF areas relative to the mean CCF area is \rmsccfarea.

\subsection{Data products}
\label{sec:data_products}
We provide CSV files with tables of: 
(1) individual observations (\texttt{rvs.csv}), 
(2) binned RV observations (with all constituent observations satisfying the filters above) (\texttt{rvs\_binned\_5.csv} \& \texttt{rvs\_binned\_10.csv}), and 
(3) daily averaged RV observations (\texttt{rvs\_daily.csv)}.
Each table includes both a combined RV based on our chosen order weights and its formal uncertainty and order RVs (e.g., rv\_NNN\_gauss) and uncertainties ($\sigma$rv\_NNN\_gauss), where NNN refers to the physical echelle order (equal to NNN=174-i, where i is the 1-based array index of orders in the NEID DRP FITS files).   
Unless otherwise specified, we report combined RVs using the order weights recorded in the  (\texttt{order\_weights\_interday}) column of the \texttt{order\_weights.csv} file.

The table of individual observations includes the name of the input L2 FITS file, several fields propagated from the FITS headers, the exposure meter and pyrheliometer data necessary to apply the filters described in \S\ref{sec:filter_weather} \& \S\ref{sec:filter_pointing}, potentially using alternative thresholds, and estimated solar-specific corrections (\S\ref{sec:solar_corrections}).  

The table of daily averaged RVs also includes statistics of that day's observations, including the number of binned observations that satisfy the above filters, the best-fit linear slope to that day's observations, the RMS deviation of that day's RVs from either a constant or linear model, and the best-fit linear slope of order RVs versus order index.  
We also provide a FITS file with daily averaged CCFs (\texttt{ccfs\_daily.fits}).

Finally, we provide several additional files with ancillary data to facilitate reanalyses of the data that would like to replace one step of the pipeline, while preserving other steps.  These include:
(1) daily extinction model parameters  (\texttt{extinction\_daily.csv}), 
(2) classical spectroscopic activity indicators (\texttt{activity.csv}) computed by NEID DRP 1.3 pipeline and copied from FITS headers for convenience and their daily averages (\texttt{activity\_daily.csv}),  and 
(3) a list of input FITS filenames with their dates and bin they were assigned to, so others can easily reproduce our exact binning of data (\texttt{filenames.csv}), and
(4) a list of observing ``runs'', their start and stop dates, RV offset adopted based on reliable LFC-calibrated orders, and offsets for each order included in our analysis (\texttt{run\_offsets.csv}).
The data release is archived on Zenodo \citep{ford_2024_neid_solar_data}.



\section{Results}
\label{sec:results}
\begin{figure}
    \centering
    \includegraphics[width=0.5\linewidth]{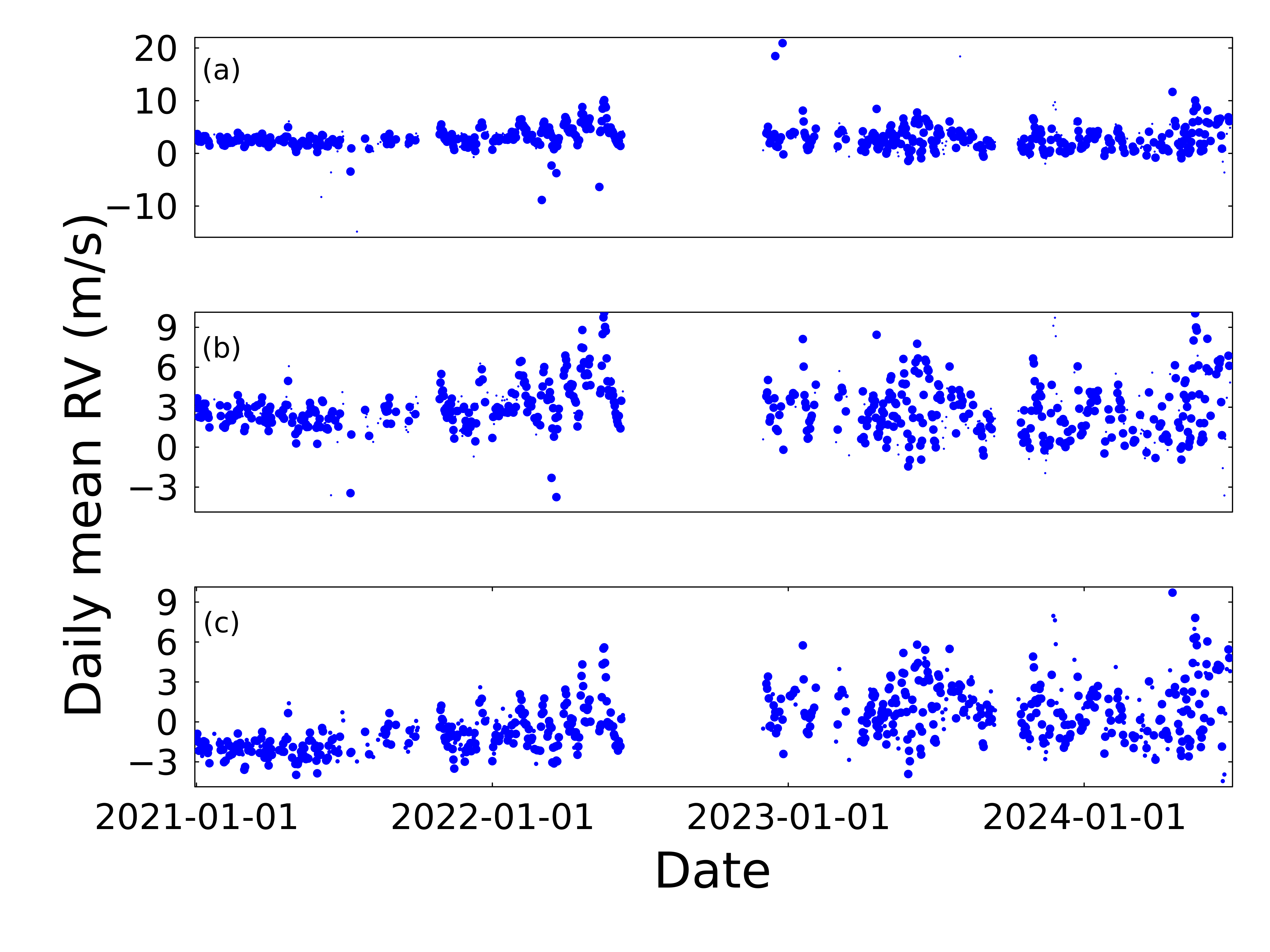}
    \caption{Top and Middle: Mean daily averaged RVs from the NEID DRP v1.3.  The two panels differ only in their y-axis scale. Bottom:  Mean daily RVs computed for this study using the Penn State Research Pipeline.  The RMS of daily average RVs is \stdmeandailyrv~ \mps.  The formal photon noise measurement uncertainties are negligible.  Both timeseries were filtered as described in \S\ref{sec:filter}.  In both panels, the small points are days with less than 15 binned observations which are more affected by solar oscillations, have lower effective signal-to-noise, and are more susceptible to including an observation contaminated by atmospheric fluctuations. \label{fig:daily_rv_vs_date_compare_drp} }
    
\end{figure}

We performed a visual inspection of the binned order RVs within each day with at least 15 binned observations to search for evidence of any additional data quality issues.  
In order to evaluate results for orders that extend beyond the ESPRESSO mask, we show results based on our custom line list.  
We compare the mean daily RV using the PSRP to those from the NEID DRP v1.3 in Figure \ref{fig:daily_rv_vs_date_compare_drp}.  For orders covered by  both masks, results were very similar with the ESPRESSO G2 mask often appearing to perform slightly better.  We leave an exploration and tuning of parameters with RvLineList for future research.

\subsection{Daily trend}
\label{sec:daily_slope}
\begin{figure}
    \centering
    \includegraphics[width=0.5\linewidth]{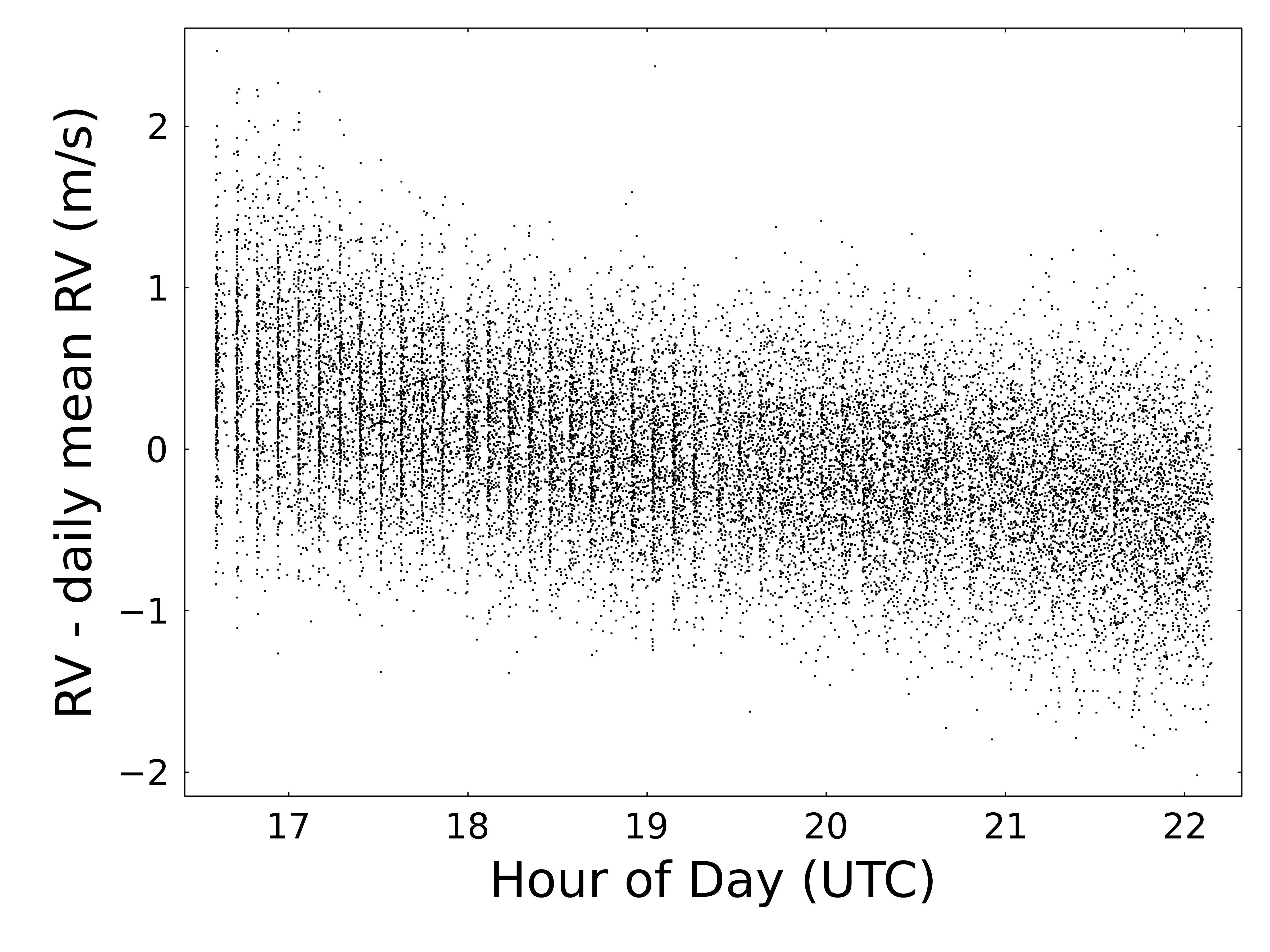} 
    \caption{We show all binned NEID RVs from the PSRP as a function of the hour of day (UTC).  The slope  suggests that the NEID wavelength calibration model has room for further improvement.}
    \label{fig:daily_rv_vs_tod}
\end{figure}

One concerning feature is an apparent slope in RVs versus time of day, as shown in Fig.\ \ref{fig:daily_rv_vs_tod}.
We find a median (over all days with sufficient data) slope of $\simeq -0.24$\mpsphr. 
The implied change in velocity over the course of a day of solar observations (over 1\mps) is significantly larger than the corrections expected based on differential extinction ($\simeq 0.06$\mps).  
Further, the steady downward trend is distinct from what is expected based on differential extinction, since that depends on airmass and would lead to a reversal in the trend each solar noon.  
These observations suggest that the trend is due to an imperfection in the daily wavelength drift model.  
Next, we look for trends in the daily RV slope versus time with either the echelle order and/or time of year to gain insights into the cause and develop an empirical model for correcting RVs of this signal.  

\subsubsection{Order}
\label{sec:daily_slope_vs_order}
\begin{figure}
    \centering
    \includegraphics[width=0.48\linewidth]{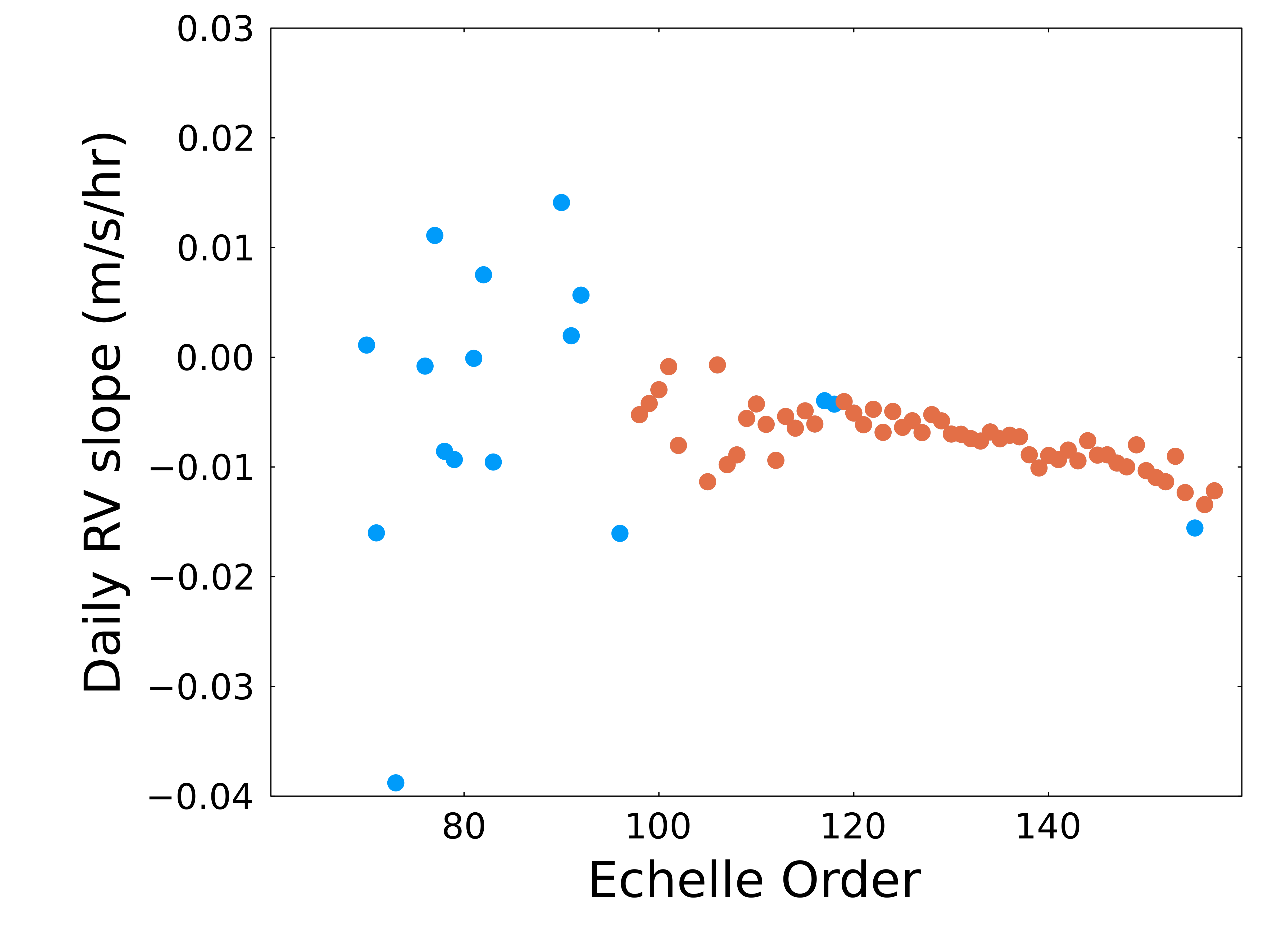}
    \includegraphics[width=0.48\linewidth]{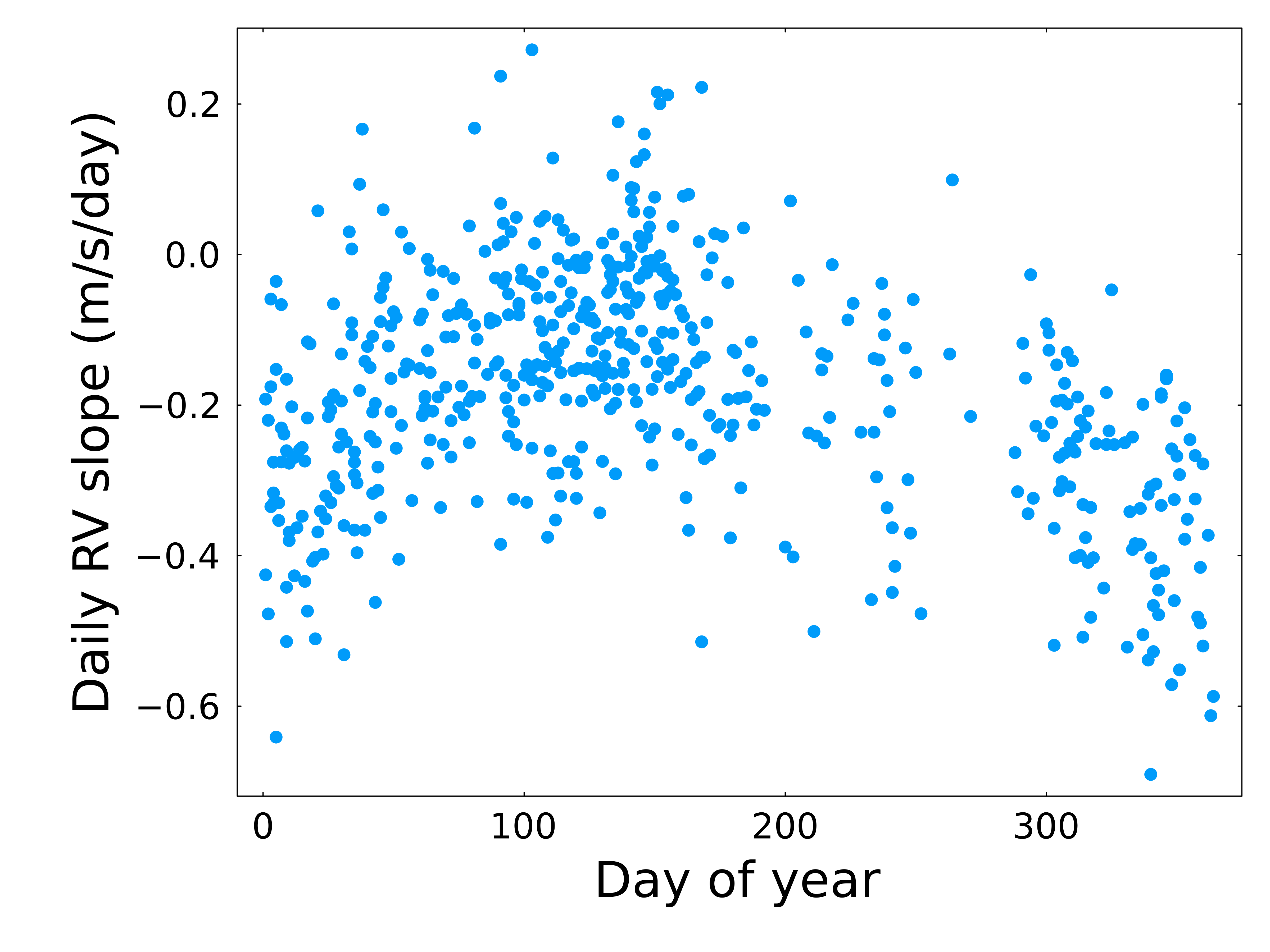}
    \caption{Left: Mean daily RV slope as a function of echelle order.  Points in red were used for fitting the trends with RV slope on order and day of year.  Right:  Mean daily RV slope as a function of the day of the year.  \label{fig:daily_slope_vs_order} \label{fig:daily_slope_vs_day}}
    
\end{figure}

The day-to-day dispersion in best-fit RV slope varies with echelle order, as shown in Fig.\ \ref{fig:daily_slope_vs_order} (left).  
After excluding several orders in the red with especially large dispersion of daily slopes (presumably due to variable absorption from water lines), the typical dispersion in daily slopes is $\simeq$ \rvslopevstimerms \mpsphr.
We fit a linear model to the daily slope versus echelle order number (using physical orders 98:101, 108:111, 119:120, 123:124, 126:143, and 145:150) and find a slope of \rvslopevstimeslope \mpsphrporder.

\subsubsection{Time of year}
\label{sec:daily_slope_vs_doy}
Next, we show the daily RV slopes versus time of year in Fig.\ \ref{fig:daily_slope_vs_day}.
Here we have computed an average slope over the orders included in the fit in \S\ref{sec:daily_slope_vs_order}.
There appears to be an annual pattern.
When fitting a sinusoidal model with (with argument $2\pi\times \mathrm{(day\ of\ year)}/365$), 
we measure an amplitudes of \amplitudesinjointslopemodel \mpsphr and \amplitudecosjointslopemodel \mpsphr for the sine and cosine components, or \amplitudetotaljointslopemodel \mpsphr for their quadrature sum.

Tables of RVs include the measured RV (\texttt{rv} and a ``corrected'' RV (\texttt{rv\_corrected}) that subtracts off an estimate of the spurious RV due to these effects.  
We adopt a linear model for the spurious RV within each day, where the slope versus time that depends linearly on both the echelle order and the sine and cosine of the annual phase. 

\subsection{Intraday RMS RV}
\label{sec:rms_rv_within_day}
\begin{figure}
    \centering
    \includegraphics[width=0.48\linewidth]{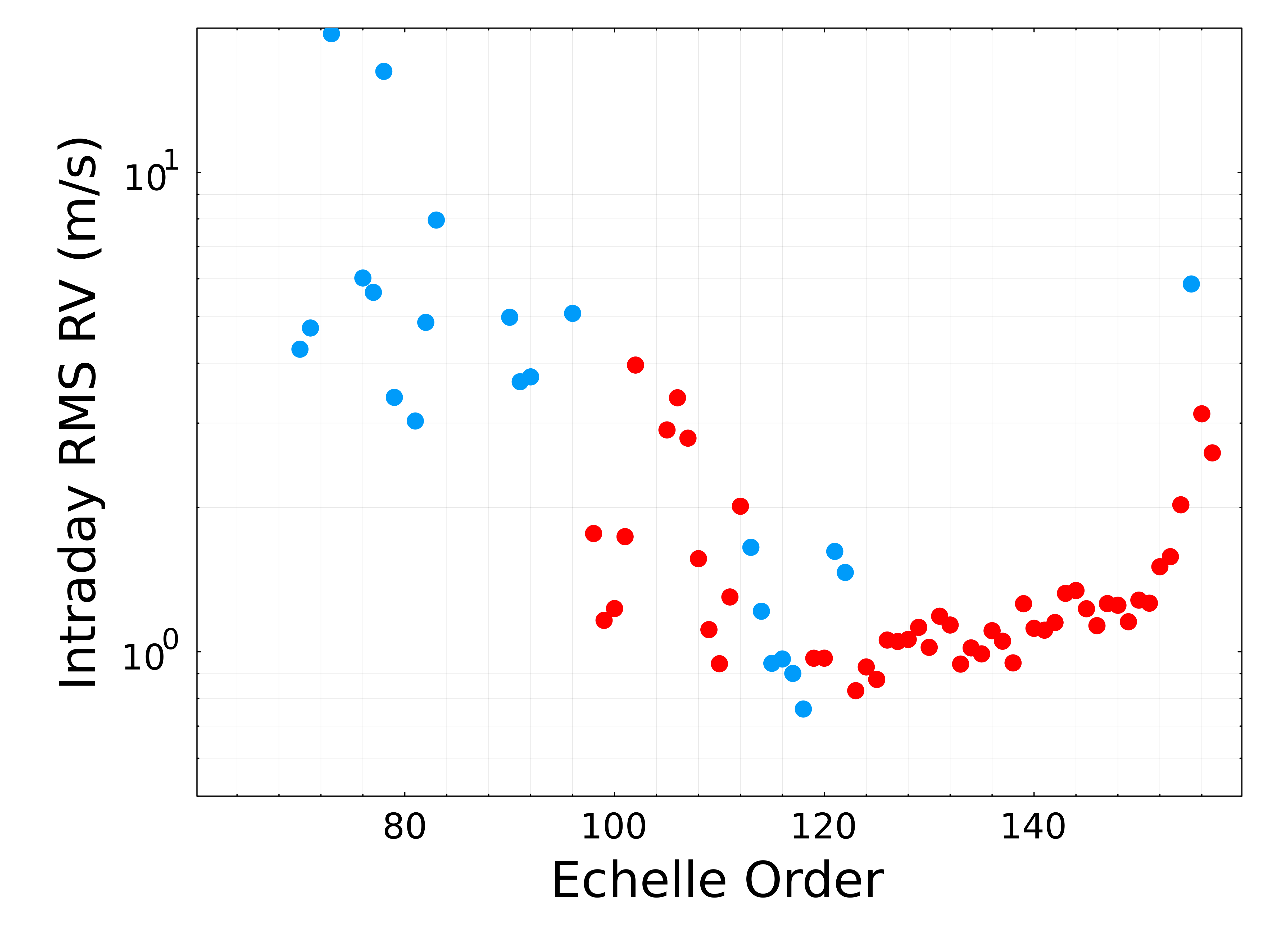}
    \includegraphics[width=0.48\linewidth]{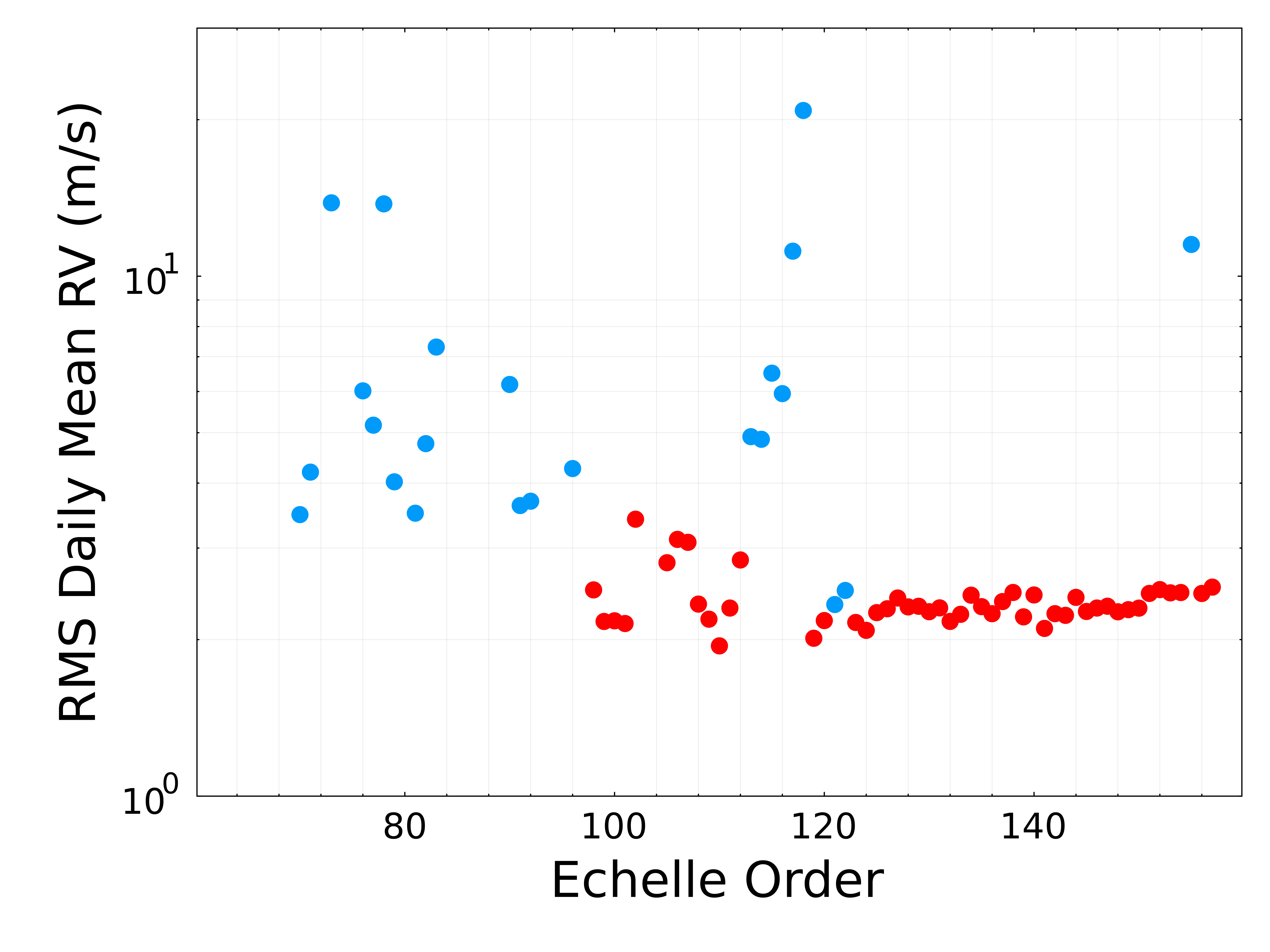}
    \caption{Left: Intraday RMS RV vs echelle order.  The RMS RV within each day was used for setting initial order weights.  Right:  Interday RMS RV vs echelle order.  These were used for setting the order weights used for comparing measurements across multiple days.  The excursions near order 115-118 are due to days where the LFC flux was low.  Including these orders in the computation of RVs results in some days showing up as large outliers.  Setting the order weights of these to zero results in significantly reduced RMS daily RVs. \label{fig:intraday_rms_rv_vs_order} \label{fig:interday_rms_rv_vs_order}}
\end{figure}

\begin{figure}
    \centering
    \includegraphics[width=0.5\linewidth]{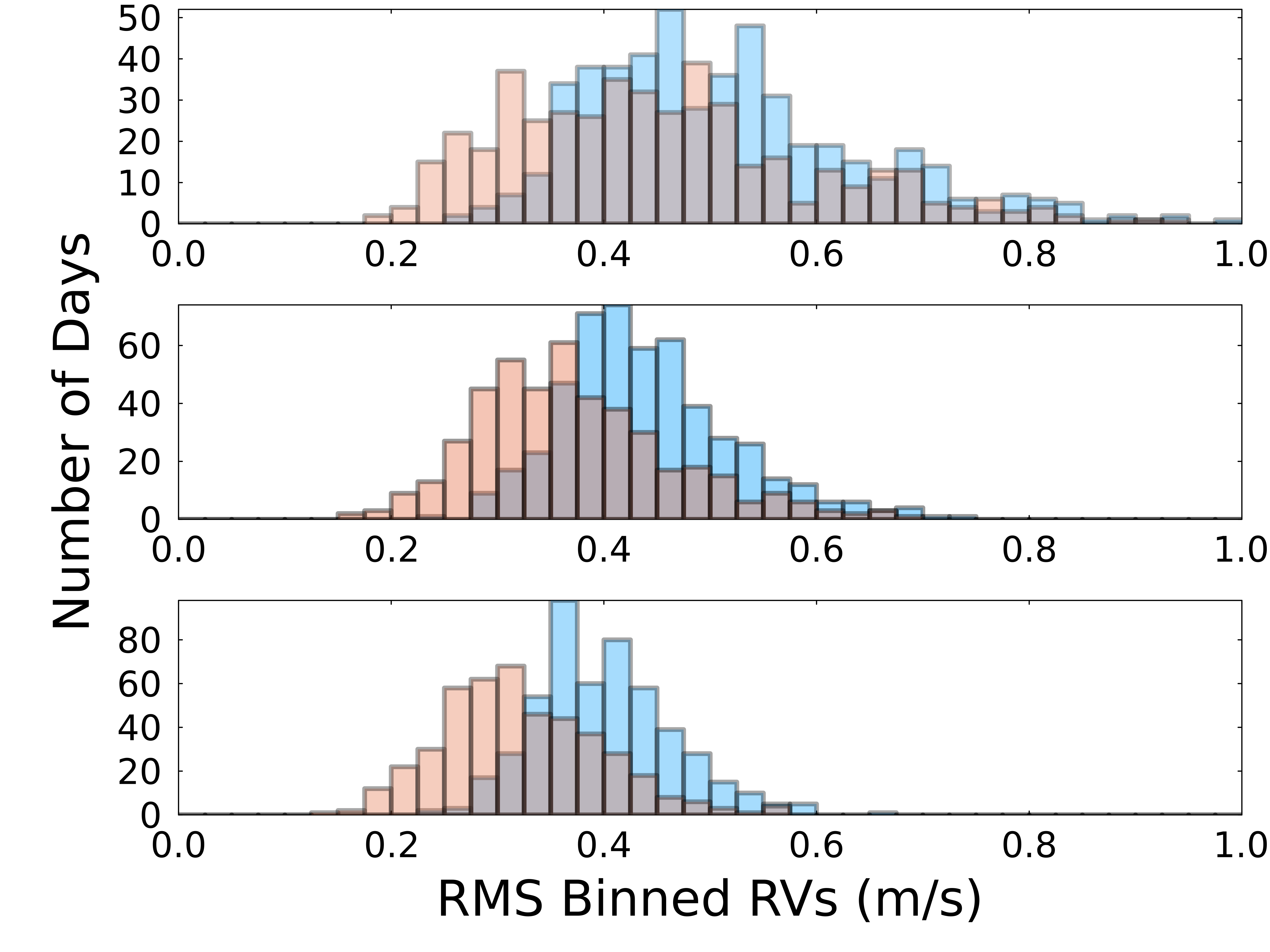}
    \caption{Histograms of RMS of binned RVs within each day.  The light blue histograms are based on binning 5 consecutive observations and at least 20 binned observations within the day.  The light red histograms are based on binning 10 consecutive observations and at least 10 binned observations within the day.   The overlpa region appears as grey.  Top:  The RMS RV is measured relative to a constant mean for each day before applying corrections for differential extinction or imperfections in the wavelength drift model.  Middle:  The RMS RV is measured relative to a constant mean for each day using RVs that include our correction for differential extinction.  This is likely to overestimate the true RMS RV due to imperfect models for differential extinction and drift in the the wavelength model. Bottom:  The RMS RV is measured relative to a best-fit linear model for the RVs within each day.  This may underestimate the RMS RV due to solar variability, since the linear model could absorb a linear trend due to changes in the solar spectrum over the course of the day.  }
    \label{fig:intraday_rms_histo}
\end{figure}

We compute the intraday RMS RV deviation from the above linear model 
for each echelle order, and report the mean over all days with at least 20 binned observations (see Figure \ref{fig:intraday_rms_rv_vs_order}, left).  
These are used to set the order weights for examining the intraday RMS RV.   
Figure \ref{fig:intraday_rms_histo} shows the distribution of the RMS deviation of the binned RV from both a constant (before correcting for differential extinction; top) and from one of two linear trends (after correcting for differential extinction).
The middle panel subtracts off the linear trend estimated based on \S\ref{sec:daily_slope}.  
The bottom panel subtracts off the best-fit linear trend for each day.  
While fitting for a linear trend within each day is intended to mitigate presumed wavelength calibration issues during solar observations, we recognize that it could also remove some astrophysical signals.  
Therefore, the distribution in the lower panel is expected to underestimate the within day RMS RV achievable with NEID.  
In contrast, the distribution in the middle panel is expected to overestimate the within day RMS RV that could be obtained with further pipeline improvements due to imperfections in the differential extinction and wavelength drift models.  
The median RMS RV when binning ten observations (\rmsjointslopemodel\mps) is slightly larger than the formal measurement uncertainties (typically 0.285\mps for a single observations) due to a combination of stellar variability (primarily pulsations and granulation, but also small scale magnetic activity), and instrumental noise sources (e.g., errors in the wavelength calibration).  

\subsection{Across Day RMS RV}
\label{sec:rms_rv_across_day}
For analyzing variations across days, we compute daily average RVs for each day with at least 10 binned observations.
Order weights are based on the inverse variance of daily averaged RVs (see Figure \ref{fig:intraday_rms_rv_vs_order} right).  
Orders with an RMS daily averaged RV greater than 3 \mps~ were assigned zero weight (and plotted with blue points).
When computing the RMS RV within a run, each (order, run) pair gets its own RV offset.  
When computing the RMS RV over the full timespan of observations, the offset between the mean RV of each run is set based on the weighted mean of the reliable LFC orders.  
For computing the final daily average RVs, we update the order weights to be proportional to the inverse of RMS of daily averaged RVs (see Figure \ref{fig:interday_rms_rv_vs_order}, right).

\subsection{Daily Average RV}
Figure \ref{fig:daily_mean_rv_vs_date} shows the mean daily RV versus date. 
The top two panels make up run 1 (pre-fire) and the bottom two panels make up run 2 (post-fire).  
Small points show the daily average RV on days with less than 15 binned observations.  
During early 2021 (top panel), the sun was still in a phase of relatively low activity.  By late 2022, we can clearly see an astrophysical signal in the RV time series on the solar rotation timescale of $\sim$26 days.

\subsection{Daily Average of Classical Spectroscopic Activity Indicators}
\label{sec:obs_classic_act_ind}
\begin{figure}
    \centering
    \includegraphics[width=0.96\linewidth]{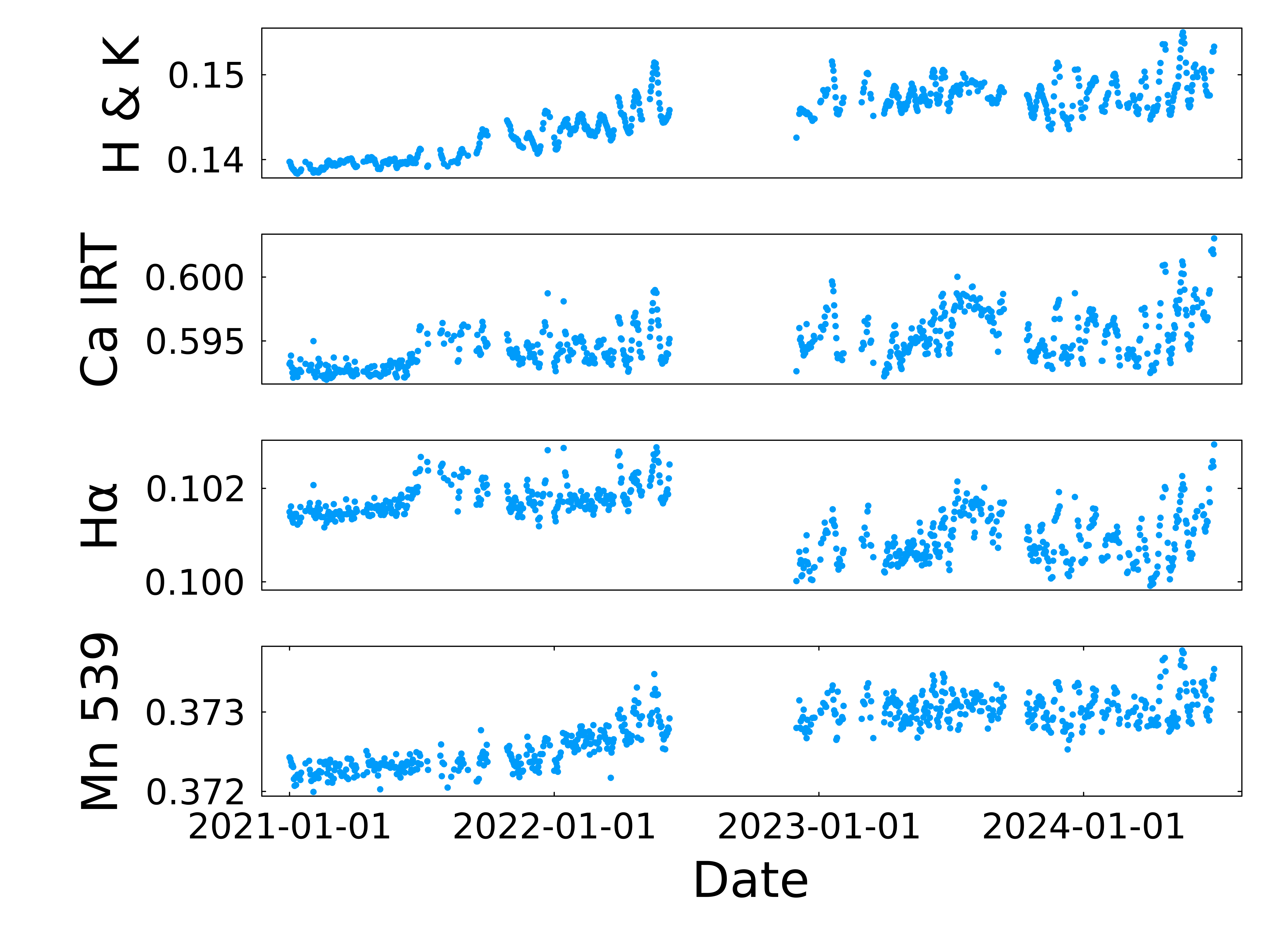}
    \caption{Daily averages of selected classical spectroscopic activity indicators versus time.  Some indicators display a substantial offset between run 1 and run 2 (large gap in late 2022) due to a pipeline configuration issue. \label{fig:activity_vs_time} }
\end{figure}

Stellar astronomers have identified several spectroscopic indicators useful for studying magnetic variability and other astrophysics.  
For Sun-like stars, the Ca II H \& K lines are particularly effective at diagnosing both rotationally modulated magnetic activity and long-term magnetic activity cycles.  
Since the Ca II H \& K lines are in the near UV, measuring this indicator is challenging for cooler stars and for some optical spectrographs.  
In these cases, other indicators such as H$\alpha$ have also proved useful for recognizing stellar magnetic activity. 
The NEID DRP provides measurements of these as well as several other classical spectroscopic indicators (He I D3 at 587.562 nm, Na I D1/D2 at 589.592 and 588.995 nm, H-$\alpha$, Ca I at 652.2795 nm, Ca II Infra-red Triplet (IRT) at 849.8018, 854.214 and 866.214nm, Na I NIR Doublet at 818.326 and 819.482, Paschen $\Delta$, and Mn I at 539.4677 nm).  
We show the time series for daily averages of four selected indicators in Figure \ref{fig:activity_vs_time}, Ca II H \& K \citep{CaIIHK}, Ca IRT \citep{Serval}, H-$\alpha$ \citep{Halpha}, and Mn 539 \citep{Mn539}.

\begin{figure}
    \centering
    \includegraphics[width=0.5\linewidth]{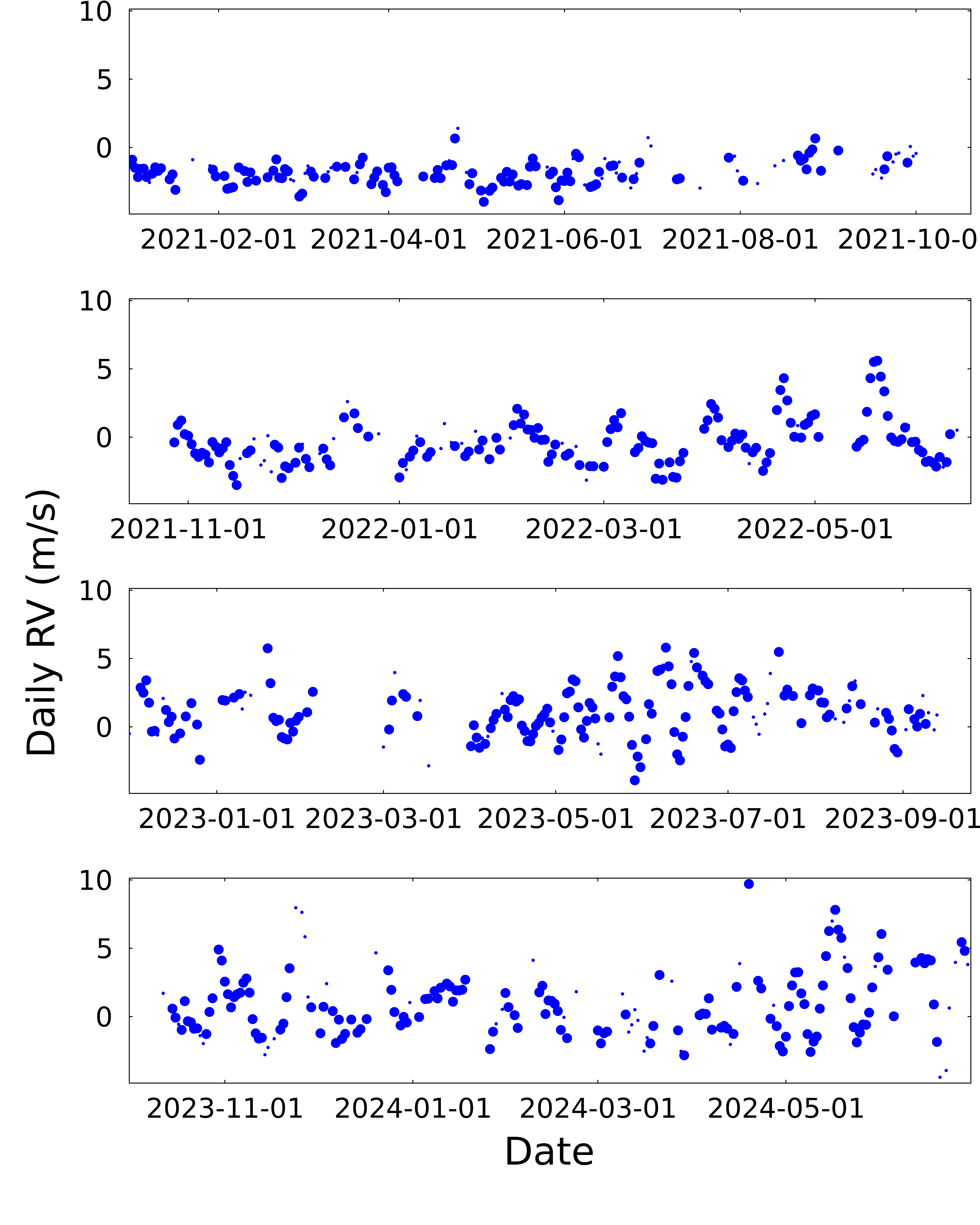}
    \caption{Mean daily RV as a function of time.  We show four sections of data each with a roughly similar number of days with observations passing our quality filters, but with precise breakpoints set by large gaps in the data.  Small dots indicate days with less than fifteen binned observations that passed our filters. \label{fig:daily_mean_rv_vs_date} }
\end{figure}

These confirm that the Sun was in a quiet state in early 2021 for the beginning of NEID solar observations and began ramping up in activity shortly before the Contreras fire shutdown.  
The Sun has been quite active since the NEID restart (for all of run 2).
The classical activity indicators demonstrate that the spurious RVs measured by the NEID Solar Telescope are dominated by astrophysical effects (rather than instrumental or pipeline), reveal periods of time with prominent rotationally-modulated activity signals, and show a long-term trend during the timespan of these NEID solar observations that qualitatively resembles that seen in the measured RVs (see Figure \ref{fig:daily_mean_rv_vs_date}). 
Thus, these measurements will prove helpful in interpreting results in \S\ref{sec:eval_mitigation} where we attempt to mitigate the effects of solar variability on measured RVs.  

\section{Evaluating Solar Variability Mitigation Strategies}
\label{sec:eval_mitigation}
While several methods have been proposed to mitigate the impact of stellar variability, it is very difficult to validate these methods on stars other than the sun, since their true center-of-mass motion is unknown.
Further, different methods can suggest significantly different ``corrections'' \citep{ZhaoEsspThree}.  
Here, we evaluate several wavelength-domain methods for mitigating stellar variability.
Each involves measuring empirical correlations between the measured RV shift and one or more input ``features'' where both are calculated from individual spectra and using those correlations to ``clean'' the measured RVs of (some of the) effects of intrinsic solar variability.
Some features were defined for other purposes and unlikely to be optimal for mitigating solar variability, but can have benefits of being easier to interpret.
Other feature sets are derived using statistical or ML methods.  
We anticipate that these are likely to have greater predictive power, but may be more difficult to interpret, to explain, and to earn the trust needed to support the discovery of Earth-analogs.  
By comparing the results of different stellar variability mitigation methods on a common high-quality solar dataset, we find many similarities and find one method that is particularly promising for future application and further development.

\subsection{Classical Spectroscopic Activity Indicators}
\label{sec:eval_classic_act_ind}
Some previous studies have attempted to clean exoplanet survey RVs by detrending with classical activity indicators like Ca II H\&K or H$\alpha$.
We apply this technique to the daily averaged solar data, so as to provide a reference point for evaluating the effectiveness of detrending against other ML-based stellar variability indicators in subsequent sections.  
We apply the same filters as for RVs and compute daily averages of the values provided for each NEID observation by the NEID DRP v1.3. 
Just as the pre/post-fire shutdown introduced offsets in the wavelength calibration, there are also clear offsets in some of the classical activity indicators. 
The bulk of these offsets have been traced to the fact that NEID DRP v1.3 did not use the blaze function derived for each observation when computed activity indicators.  
This issue should be fixed starting in DRP v1.4.  
For this study, we simply add unknown coefficients for the mean of each activity indicator during each of run 1 and 2.  
Among classical activity indicators, detrending against Ca II H \& K was the most effective at reducing the RMS of daily averaged RV from \stdmeandailyrv~ \mps to \rmsrvsCaHKII~ \mps.  
For instruments that can not measure Ca II H \& K, H-$\alpha$ is a common alternative spectroscopic indicator.  
Detrending against H-$\alpha$ reduce the RMS of daily averaged RV to \rmsrvsHalpha~ \mps.  
Previous studies have noted that often there appears to be a lag between local extrema of the RV timeseries and spectroscopic activity indicators such as Ca~II~H \& K \citep{ColierCameron2019}.
As this study focuses on purely wavelength-domain methods for mitigating solar variability, we leave a joint analysis that allows for temporal lags for a future study.

\subsection{Daily Red \& Chromatic RVs}
\label{sec:chromatic_rvs}
\begin{figure}
    \centering
    \includegraphics[width=0.5\linewidth]{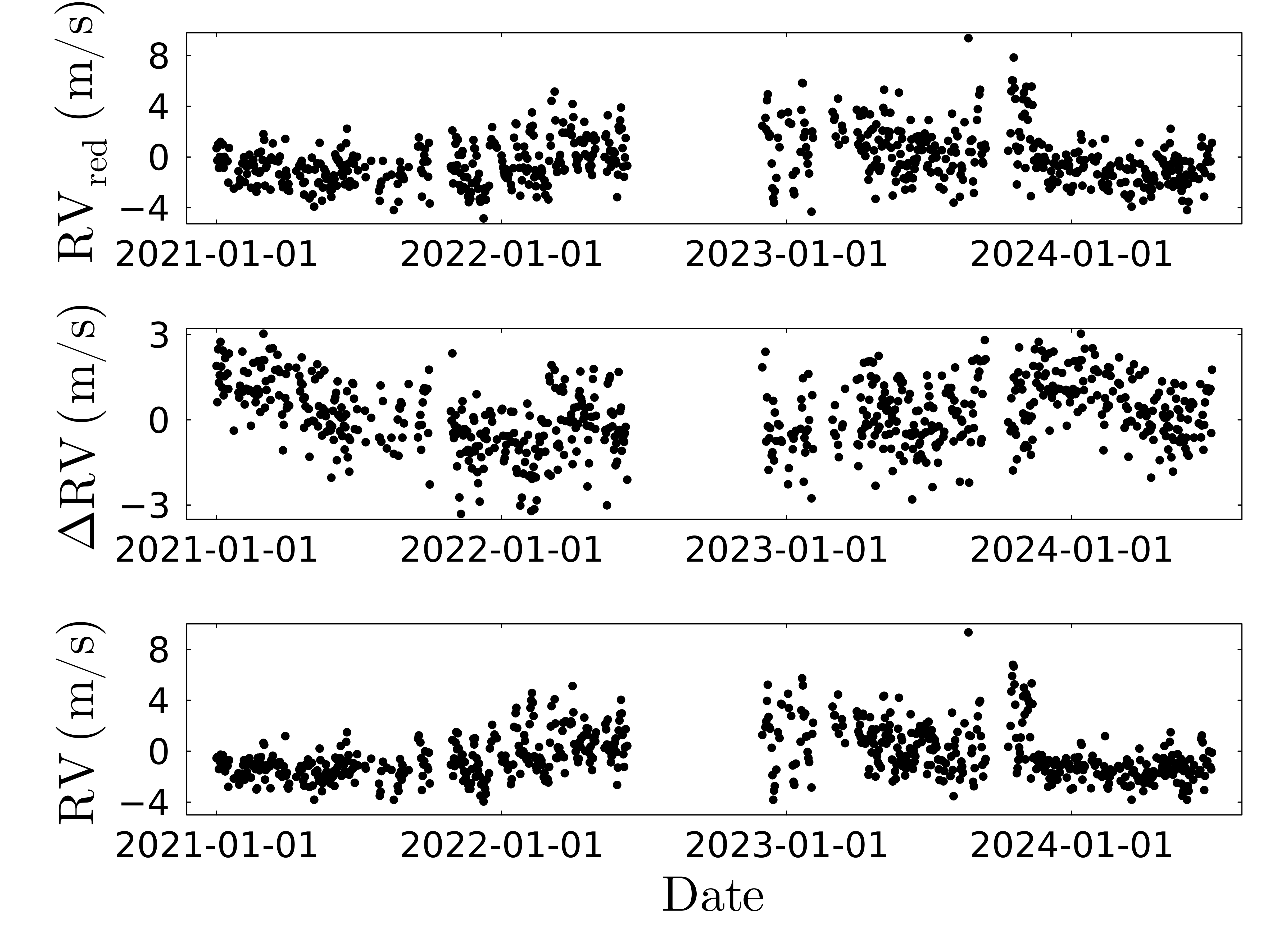}
    \caption{Top: Daily averaged Red RVs based on echelle orders 70-96 versus time.  Middle:  Daily averaged chromatic RVs (difference between the Red RVs and Optical RVs based on echelle orders $\geq~97$) versus time.  Bottom:  Daily mean of Optical RVs after decorrelating based on the chromatic RVs.}
    \label{fig:chromatic_rvs_vs_time}
\end{figure}

\begin{figure}
    \centering
    \includegraphics[width=0.48\linewidth]{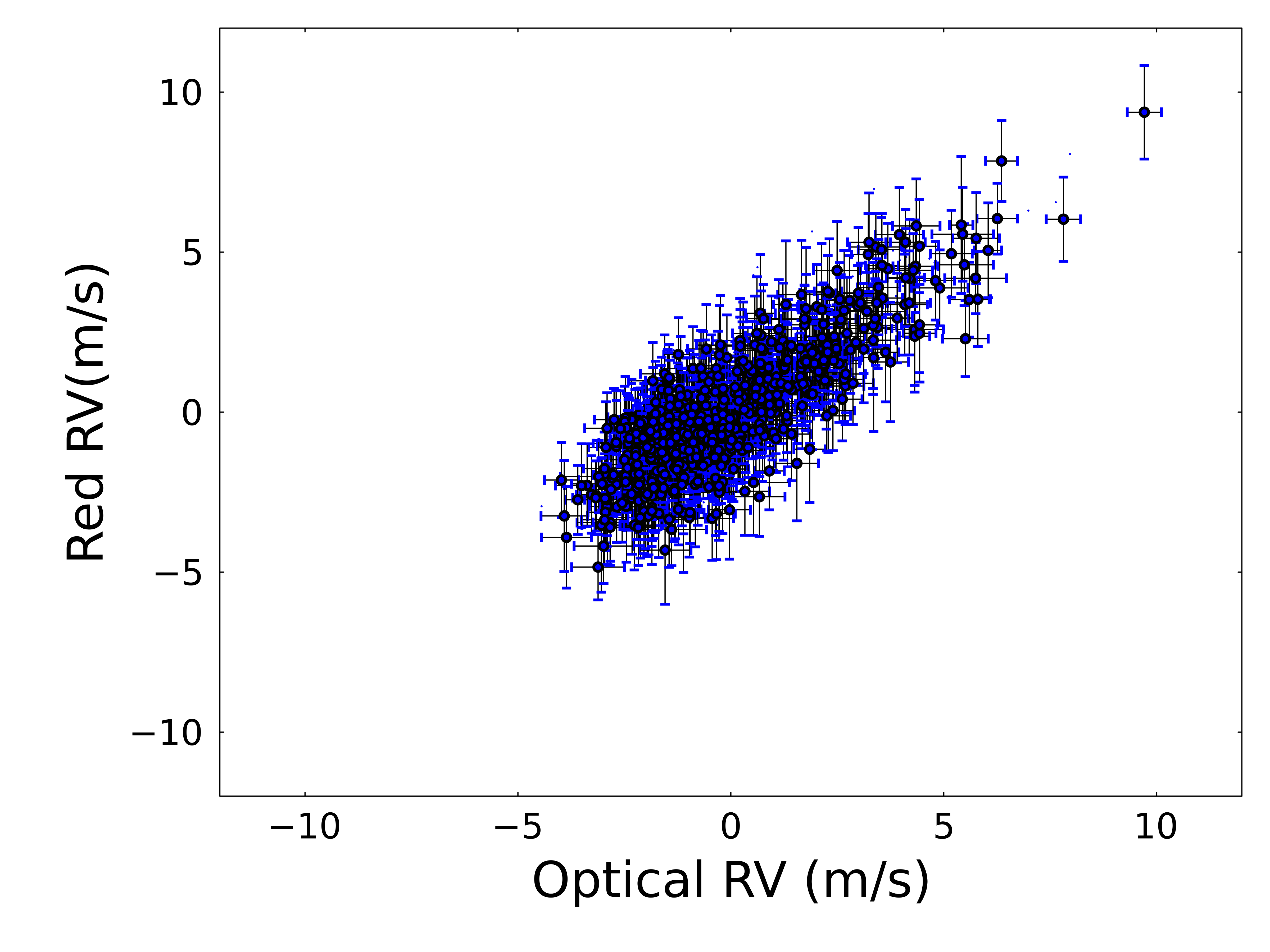}
    \includegraphics[width=0.48\linewidth]{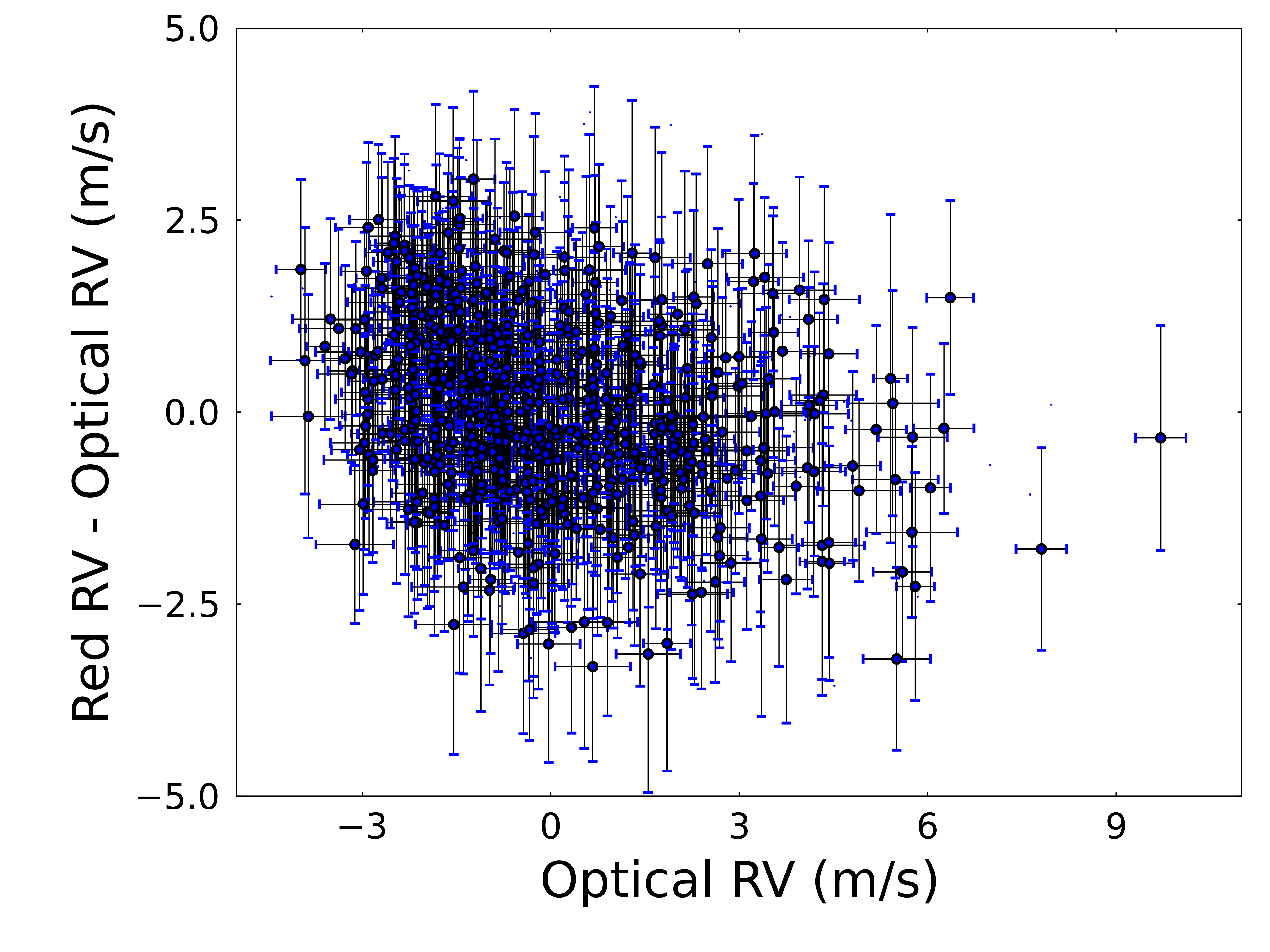}
    \caption{Left: Daily average RVs for days with at least 15 binned RVs using our baseline order weights (``optical'') and weights using only ``red'' portions of the spectrum.  
    Errorbars show the within-day RMS RV for based on the two sets of order weights.
    Right: Difference in Red RV and Optical RV versus Optical RV. \label{fig:nir_vs_optical}}
\end{figure}

One feature of NEID is its broad wavelength grasp.  
This was intended to allow NEID to measure classical stellar activity indicators that run from near UV/blue to far red/NIR.
It also has the potential to be useful for measuring differences in apparent RV with wavelength or order. 
In Figure \ref{fig:chromatic_rvs_vs_time}, we examine this wavelength dependence by dividing the order RVs into two disjoint groups that we refer to as Optical RVs (physical echelle orders $\geq~97$ or $\lambda\le631.4\mathrm{nm}$) and Red RVs (echelle orders 70-96, $632.4\mathrm{nm} \le \lambda < 884.0\mathrm{nm}$). These are daily average RVs, with order weights based on our standard order weighting.
There is a strong correlation between the optical and Red RVs, and a moderate correlation between the optical RVs and the difference between the two (``chromatic RV'') (see Figure \ref{fig:nir_vs_optical}).
Attempting linear decorrelation using the chromatic RV reduces the RMS of daily mean RVs from \stdmeandailyrv~ \mps~ to \stdmeandailyrvchromaticcorrectedred~ \mps.  
The chromatic RVs appear to be tracing long-term changes more than stellar variability on the rotational timescale. 

\subsection{CCF-based Mitigation Strategies} 
\label{sec:eval_ccf}
Next, we attempt to mitigate the effects of stellar variability on the daily mean RVs by considering several proposed stellar variability indicators calculated from the CCF.
We compute a mean daily CCF for \numdaysmeanslope~ days with at least 15 binned observations passing our data quality filters.
Below, we apply several approaches for characterizing the CCF shape and test their effectiveness for linear decorrelation with measured solar RVs.  

\subsubsection{Classical CCF Shape Indicators} 
\label{sec:eval_ccf_shape_classical}
\begin{figure}
    \centering
    \includegraphics[width=0.48\linewidth]{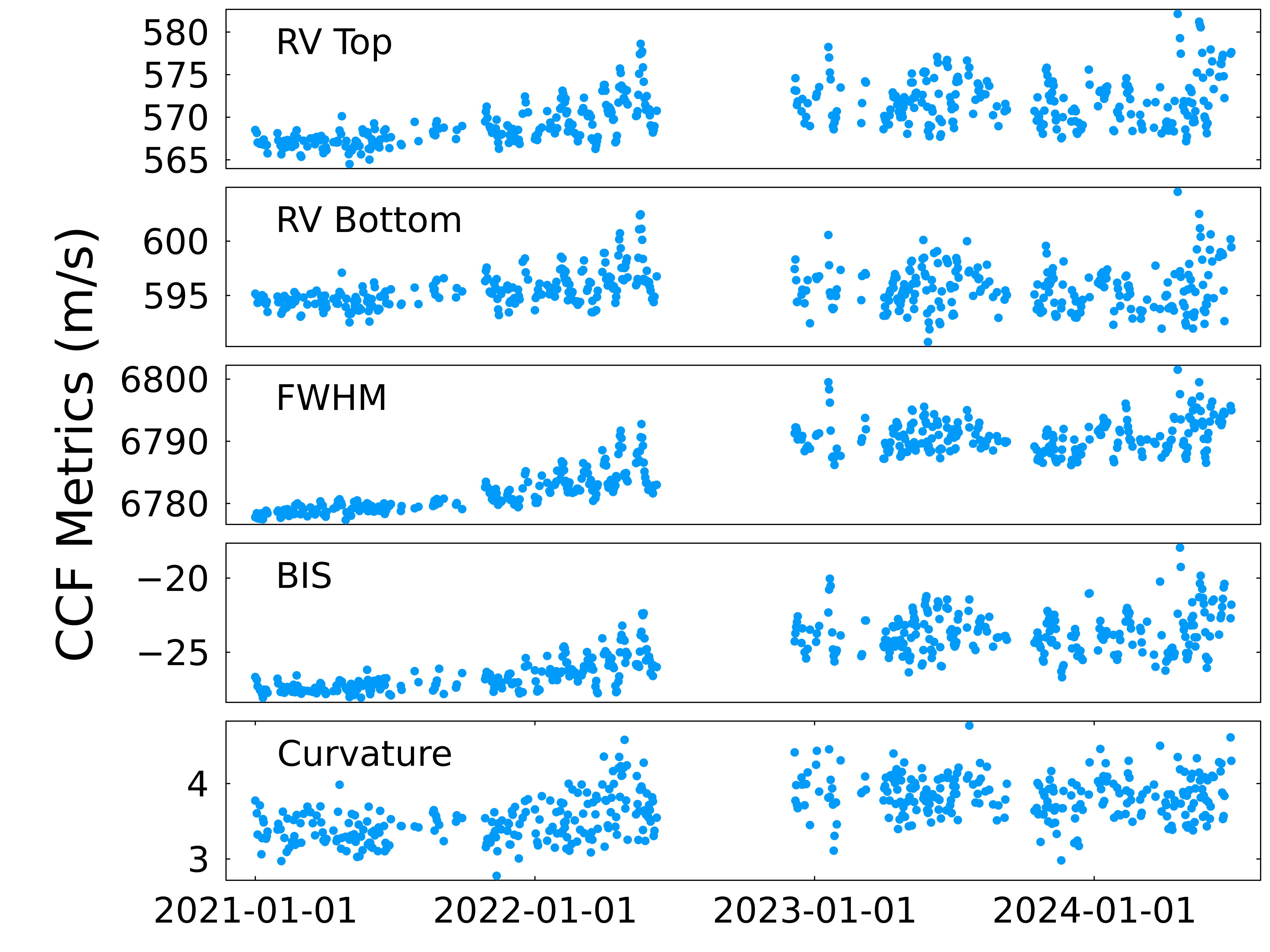}
    \includegraphics[width=0.48\linewidth]{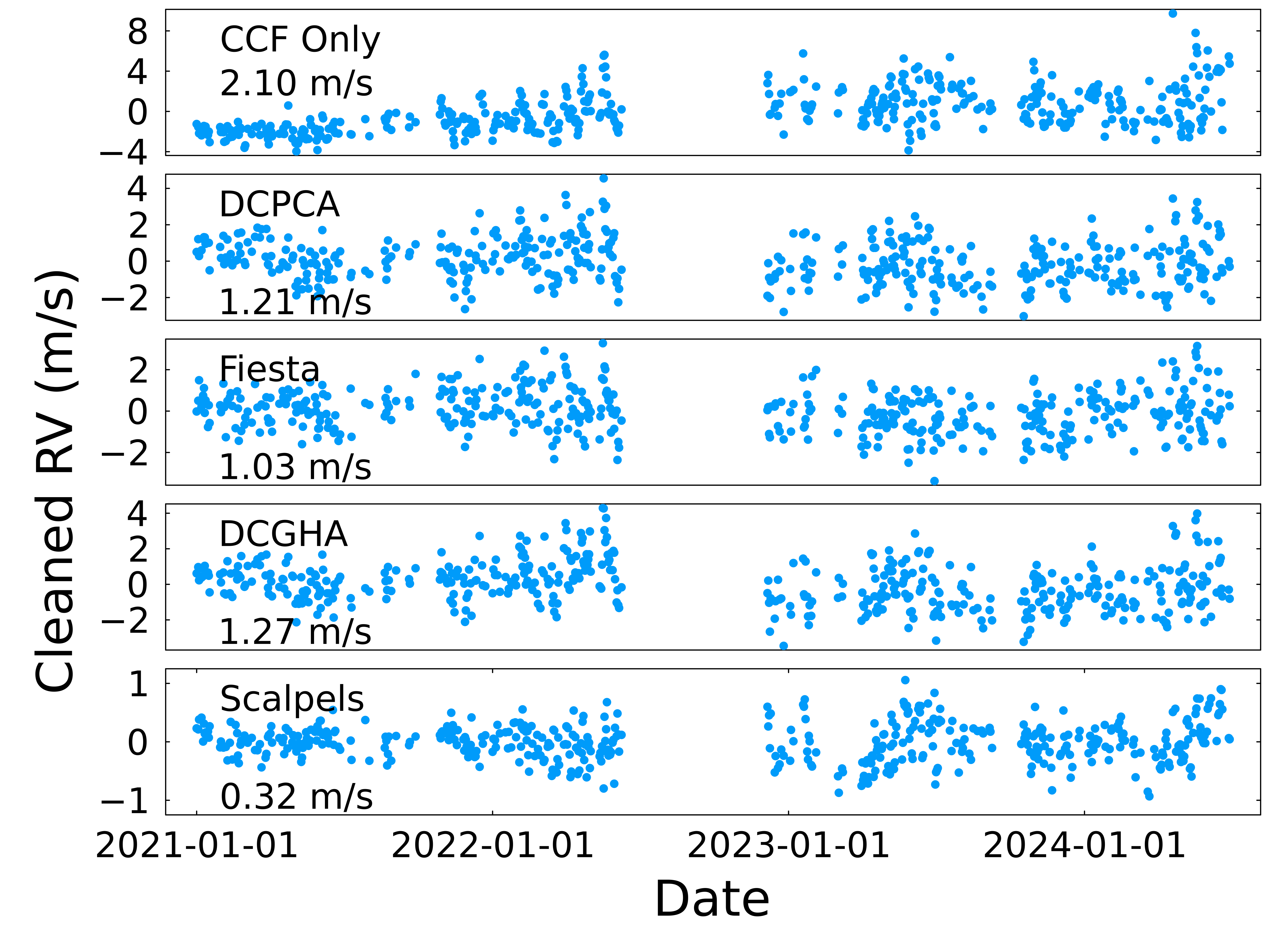}
    \caption{Left:  Daily-averaged CCF-based solar variability metrics.  
    Since the CCF area is very nearly constant, the product of the CCF FWHM and CCF contrast (ratio of depth to CCF normalization) is very nearly constant, and the CCF contrast (not shown) would appear like an inverted version of the CCF FWHM.  
    Right:  Daily-averaged RVs after detrending with different solar variability indicators. }
    \label{fig:ccf_metrics_vs_time} 
\end{figure}

Figure \ref{fig:ccf_metrics_vs_time} (left) shows time series for multiple common daily CCF shape measurements, including the full width at half maximum (FWHM; measured based on fitting a Gaussian line profile to the CCF), the bisector inverse slope (BIS),
and the bisector curvature (see \S\ref{sec:methods_biss}).
The CCF FWHM and BIS show little variability in 2021, but significant rotational-linked variability in subsequent years, and a long term increase which we attribute to the solar magnetic cycle.  
Both are consistent with trends observed in Ca II H\&K and other classical activity indicators.  
When we attempt to detrend using any either the classical CCF shape indicators, we find using  the CCF BIS yields the lowest RMS of daily RVs, \rmsrvsbis~ \mps.  
While an improvement, this proved less effective than using Ca II H\&K in \S\ref{sec:eval_classic_act_ind}. 

\subsubsection{Doppler-constrained Principal Components Analysis (DCPCA)} 
\label{sec:eval_ccf_dcpca}
Next, we consider using several different features measured from daily-averaged CCFs using data-driven or machine learning (ML) approaches.  
First, we consider a variant of Doppler-constrained Principal Components Analysis \citep[][DCPCA]{JonesDcpca,GilbertsonDcpca}.
Instead of analyzing the spectrum directly with DCPCA, we apply it to the daily-averaged CCFs.  
We summarize the steps: 
(1) shift CCFs to remove large known barycentric corrections, 
(2) compute a template CCF, 
(3) subtract the template from each CCF to form a residual CCF, 
(4) project the residual CCF onto the derivative of the template CCFs to estimate the RVs, 
(5) compute the projection of the residual CCF onto the plane orthogonal to the RV component ($\mathrm{CCF}_{\mathrm{shape}}$), 
(6) perform a standard principal components analysis (PCA) of $\mathrm{CCF}_{\mathrm{shape}}$'s, and
(7) perform linear regression on the PCA scores with the estimated RVs (after subtracting the Doppler shifts from any known planets) to estimate model coefficients for predicting the portion of the estimate RVs that can be attributed to shape driven RVs ($\mathrm{RV}_{\mathrm{shape}}$),
(8) predict the shape driven RVs ($\mathrm{RV}_{\mathrm{shape}}$) for each observation based on the PCA scores, and
(9) estimate the shift-driven or ``cleaned'' RVs by subtracting $\mathrm{RV}_{\mathrm{shape}}$ from the estimated RVs.

The results are shown in second right panel of Figure \ref{fig:ccf_metrics_vs_time}.
DCPCA has one important algorithmic parameter, the number of PCA feature vectors and scores to retain in steps 7 \& 8 when computing ($\mathrm{RV}_{\mathrm{shape}}$).
The blue curve in the left panel of Figure \ref{fig:rms_vs_num_features} shows how the RMS of cleaned RV decreases as the number of scores and feature vectors increase.  
Using even a single feature vector yields a substantial improvement.   
There is a gradual improvement with the use of additional feature vectors 2--8.
Some algorithms are susceptible to increasing out-of-sample prediction errors when the number of scores and feature vectors used increases since the model coefficients can start to increase due to overfitting.  
Fortunately, we find that DCPCA yields a long plateau extending to several dozen feature vectors.
In such cases, a common strategy is to pick the smallest number of feature vectors that reaches the plateau.  
Applying DCPCA to the NEID solar daily-averaged CCFs dataset yields an RMS RV of \rmsrvsdcpca \mps.  
This represents a substantial improvement over detrending RVs based on either classical spectroscopic activity indicators or traditional CCF shape metrics.

One could worry about the potential for ``overfitting'', or overestimating the improvement in RMS RVs when applied out-of-sample, due to using the same spectra being used in steps 6 \& 7 (``training'') and steps 8 \& 9 (``testing''). 
To address this concern, we applied a form of cross-validation, where we train and test using disjoint subsets of daily-averaged CCFs.
Rather than drawing random subsets of our data, we consider two physically motivated partitionings.
Our primary cross validation test is based on separating the data taken as part of run 1 and run 2, i.e., before and after the Contreras fire.  
Between the two runs the instrument was warmed up to its long-term safe mode and cooled back down, and we need to allow for RV offset between runs in any analysis. 
There is a multi-month gap between the observations, so there is negligible risk of information about the distribution of magnetic active regions during the training set providing useful information for the test set.
The results are shown in the blue curve in the right panel of Figure \ref{fig:rms_vs_num_features}.  
The solid (dotted) line corresponds to training of run 2 (1) and testing on run 1 (2).  
Since there are clear differences in the level of solar activity between the two runs, the two train/test pairs are expected to yield different results.
For these reasons, our train/test scenario represents a much steeper  challenge than a more typical application of cross validation. 
When using five DCPCA scores, we find out-of-sample RMS's of the cleaned daily RVs of  \cvrmsrvsdcpcafivepre~ and \cvrmsrvsdcpcafivepost~ \mps, only slightly larger than when we trained and tested on the full dataset.  
Using ten DCPCA scores, the  out-of-sample RMS's of the cleaned daily RVs improve to \cvrmsrvsdcpcaninepre~ and \cvrmsrvsdcpcaninepost~ \mps.  

We also performed an analogous cross validation scheme separating the data into four sections (shown in separate panels of Figure \ref{fig:daily_mean_rv_vs_date}), where we train using three segments and test using the one held-out segment.  
The performance was typically comparable or improved relative to our reference cross validation using runs 1 \& 2.  
This is not unexpected, since using four segments has the advantage of using a larger amount of training data (and to a lesser extent the effect of including a few training observations near in time to a few testing observations).  

\subsubsection{FourIEr phase SpecTrum Analysis (FIESTA)}
\label{sec:eval_ccf_phiesta}
Next, we apply the \phiesta algorithm \citep{Zhao2022Fiesta} to the daily NEID solar CCFs.
Here, we summarize the steps: 
(1) shift CCFs to remove large known barycentric corrections, 
(2) normalize each CCF by the ``CCF continuum'', 
(3) transform the normalized CCFs by negating them and adding one (i.e., $1-$CCF), 
(4) compute coefficients for discrete Fourier transform on the transformed CCFs, 
(5) estimate the RVs from each Fourier mode ($RV_{k}$'s) using the Fourier phase information,
(6) subtract the weighted mean RV from the Fourier modes to obtain the  shift-invariant $\Delta RV_k$'s,
(7) perform linear regression on the $\Delta RV_k$'s with the estimated RVs (after subtracting the Doppler shifts from any known planets) to estimate model coefficients for predicting the portion of the estimate RVs that can be attributed to shape driven RVs ($\mathrm{RV}_{\mathrm{shape}}$),
(8) predict the shape driven RVs ($\mathrm{RV}_{\mathrm{shape}}$) for each observation based on the $\Delta RV_k$'s,
(9) estimated the shift-driven or ``cleaned'' RVs by subtracting $\mathrm{RV}_{\mathrm{shape}}$ from the estimated RVs.

The results are shown in the third panel on the right of Figure \ref{fig:ccf_metrics_vs_time} and with the red curves in Figure \ref{fig:rms_vs_num_features}.
We find that \phiesta yields an RMS of daily cleaned RV of \rmsrvsfiesta \mps\ when using 6 feature vectors.
Like DCPCA, \phiesta also yields a broad plateau, but reaches this plateau with just four feature vectors.  
(We caution that the number of feature vectors that \phiesta needs depends on the velocity span of the input CCFs.  We adopted a velocity span of 25 \kmps, as we found that concentrated the useful information into a small number of scores and feature vectors.)  
When we perform cross validation using runs 1 and 2, we find that RMS of \phiesta's out-of-sample RV predictions for run 2 (\cvrmsrvsfiestafivepre \mps) are significantly worse than for run 1 (\cvrmsrvsfiestafivepost \mps). 
These are somewhat smaller than the analogous from analysis applying DCPCA to the daily averaged CCFs.

\subsubsection{Doppler-constrained Gauss-Hermite Analysis (DCGHA)} 
\label{sec:eval_ccf_gh}
Next, we apply Doppler-constrained Gauss-Hermite Analysis (DCGHA) which combines ideas from several previous studies, particularly \citet{JonesDcpca,Holzer_SAFE,GilbertsonDcpca}. 
Here, we summarize the steps: 
(1) shift CCFs to remove large known barycentric corrections, 
(2) compute a template CCF and estimate its mean and width by fitting a Gaussian line profile, 
(3) subtract the template from each CCF to form a residual CCF, 
(4) project the residual CCF onto the derivative of the template CCFs to estimate the RVs, 
(5) compute coefficients for projecting the residual CCF onto Gauss-Hermite basis functions using the mean and width estimated in step \#2, 
(6) perform linear regression on the Gauss-Hermite coefficients with the estimated RVs (after subtracting the Doppler shifts from any known planets) to estimate model coefficients for predicting the portion of the estimated RVs that can be attributed to shape driven RVs ($\mathrm{RV}_{\mathrm{shape}}$),
(8) predict the shape driven RVs ($\mathrm{RV}_{\mathrm{shape}}$) for each observation based on the Gauss-Hermite coefficients,
(9) estimate the shift-driven or ``cleaned'' RVs by subtracting $\mathrm{RV}_{\mathrm{shape}}$ from the estimated RVs.

The results are shown in the fourth panel on the right of \ref{fig:ccf_metrics_vs_time} and with the green curves in Figure \ref{fig:rms_vs_num_features}.
There is little benefit to including the first, second and third Gauss-Hermite coefficients unless the fourth Gaussian-Hermite is also included.  
(The number of feature vectors used is one greater than the order due to the 0th order coefficient which is quite useful even by itself.)  
When using 5 feature vectors, we find that the Doppler-constrained Gauss-Hermite analysis yields an RMS daily cleaned RV of \rmsrvsgausshermite~ \mps,
very similar to that of DCPCA once both use at least 6 feature vectors.  
This is not unexpected.  
Since both are linear function of the CCF, they should yield nearly identical results if the maximum number of feature vectors were used by both algorithms.  
The two algorithms are taking different approaches to choosing how to parameterize and reconstruct the projection of the CCFs onto the plane orthogonal to a Doppler shift.
Once both have retained most of the useful information, their predictions for $\mathrm{RV}_{\mathrm{shape}}$ are very similar.

When we perform cross validation using runs 1 and 2, we find that RMS of DCGHA's out-of-sample RV predictions for runs 1 (\cvrmsrvsgausshermitetenpre~ \mps) and 2 (\cvrmsrvsgausshermitetenpost~ \mps) are comparable to the results from DCPCA, once sufficient feature vectors are included.
Interestingly, the RMSs for the two test sets are more similar to each other than when using DCPCA or \phiesta.  
Based on comparing the predictive performance from analyzing the full data set and subsets for cross validation purposes, the DCGHA appears to perform better with smaller datasets than DCPCA or \phiesta.

\subsubsection{Self-Correlation Analysis of
Line Profiles for Extracting Low-amplitude Shifts (Scalpels)} 
\label{sec:eval_ccf_scalpels}
Finally, we apply Self-Correlation Analysis of
Line Profiles for Extracting Low-amplitude Shifts \citep[Scalpels;][]{CollierCameronScalpels} to the NEID daily solar CCFs . 
Here, we summarize the steps: 
(1) shift CCFs to remove large known barycentric corrections, 
(2) compute the discrete autocorrelation function (ACF) of each CCF, 
(3) perform a Singular Value Decomposition (SVD) on the autocorrelation function, 
(4) perform linear regression on the eigenvalues of the SVD decomposition  with the estimated RVs (after subtracting the Doppler shifts from any known planets) to estimate model coefficients for predicting the portion of the estimate RVs that can be attributed to shape driven RVs ($\mathrm{RV}_{\mathrm{shape}}$),
(5) predict the shape driven RVs ($\mathrm{RV}_{\mathrm{shape}}$) for each observation based on the the ACF of the CCF for that observation,
(6) estimate the shift-driven or ``cleaned'' RVs by subtracting $\mathrm{RV}_{\mathrm{shape}}$ from the estimated RVs.

\begin{figure}
    \centering
    \includegraphics[width=0.48\linewidth]{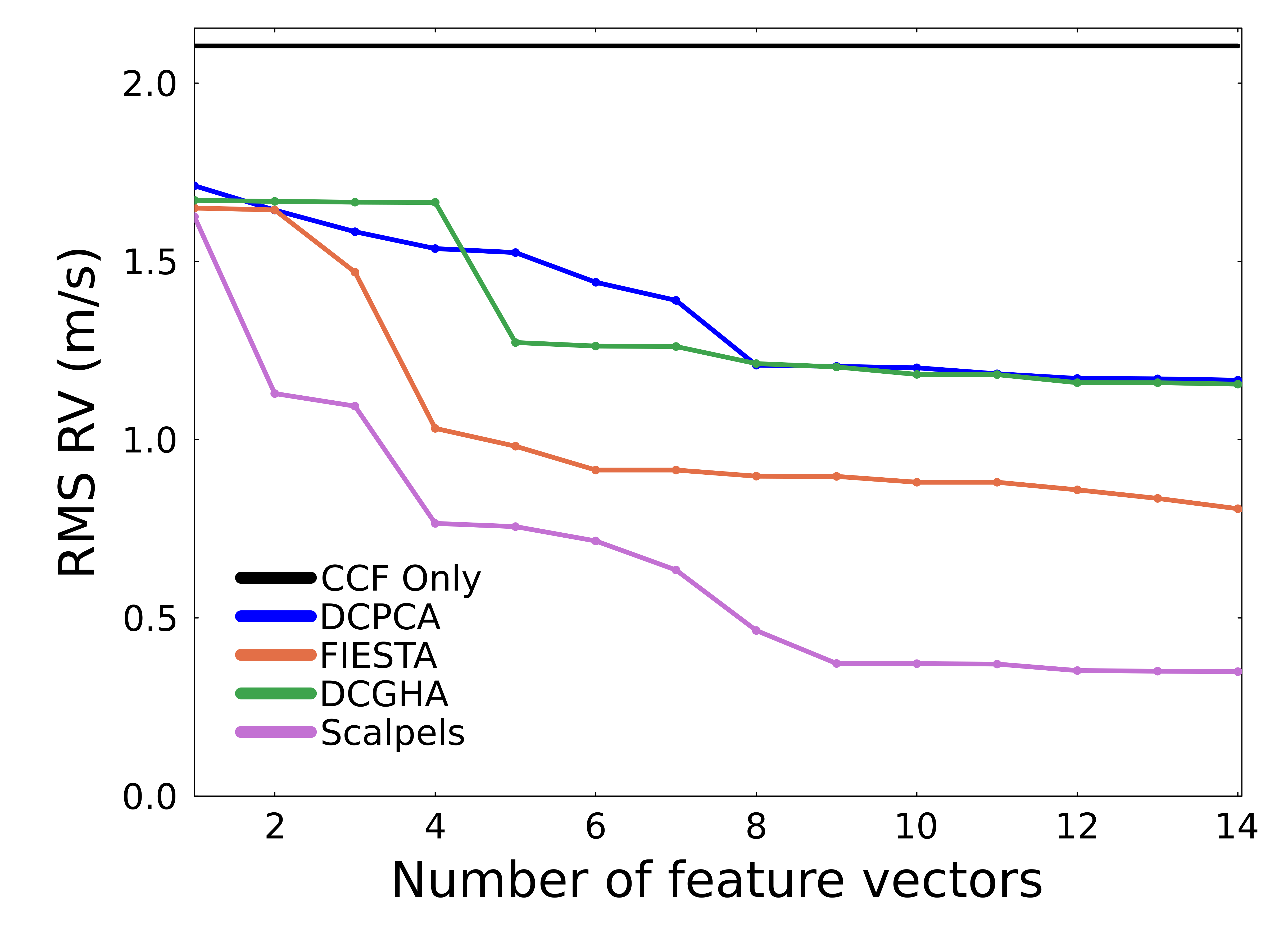}
    \includegraphics[width=0.48\linewidth]{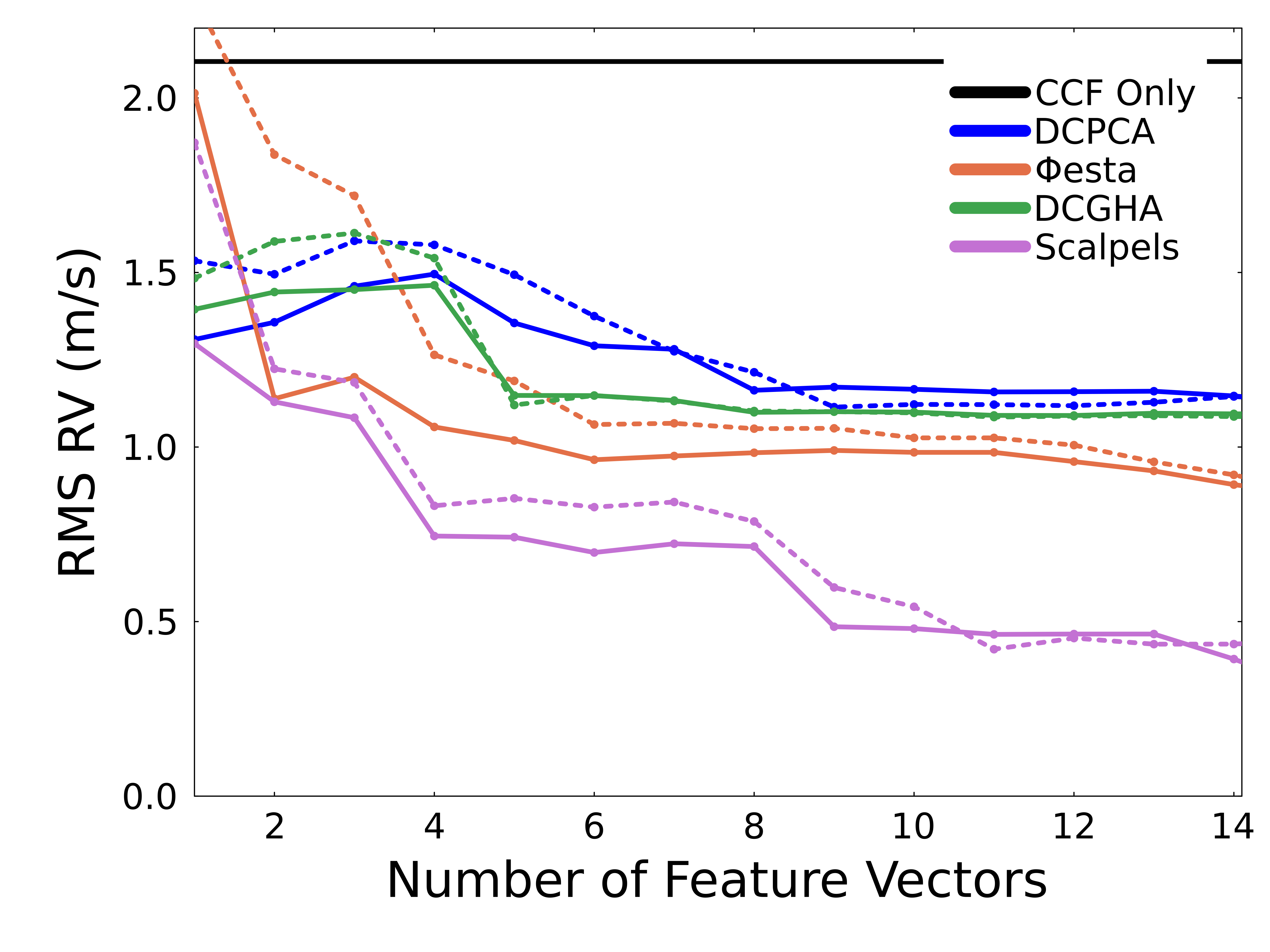}
    \caption{RMS of daily average RVs after detrending with different sets of data-driven stellar variability indicators as a function of the number of indicators used. Left: Results from training and testing each method to the full data set.  Right:  Results from training each method on one of two runs and testing on the other run.  The dotted (solid) curve show results training on run 1 (2) and testing on run 2 (1).}
    \label{fig:rms_vs_num_features}
\end{figure}

The RMS of daily RVs after cleaning with 5 Scalpels features (\rmsrvsscalpelsfive~ \mps) is a substantial improvement over the CCF-based RV and significantly less than that of any of the other algorithms considered in this study.
Allowing Scalpels to use ten features results in additional improvements (\rmsrvsscalpels \mps; see Figure \ref{fig:rms_vs_num_features} left).
Scalpels performs better than the other methods considered when holding the number of feature vectors used fixed as long as at least 2 features are used.
Scalpels shows a rapid improvement as the number of features used increases from 1 to 4, a stall from 5 to 7, and then a very broad plateau for 8 or more features used.   

When we perform cross validation using runs 1 and 2 and less than ten features (see Figure \ref{fig:rms_vs_num_features} right), we find that RMS of Scalpels's out-of-sample RV predictions for run 1 (\cvrmsrvsscalpelsfivepost~ \mps) are less than those for run 2 (\cvrmsrvsscalpelsfivepre~ \mps, values for 5 feature vectors).
When training on either run 1 or 2 alone, more feature vectors are needed before it reaches its plateau.  
When we performed cross validation scheme training on three of the four sections (shown in separate panels of Figure \ref{fig:daily_mean_rv_vs_date}) and testing on the held-out segment, Scalpels performs better than when using less training data, acheiving a smaller RMS of 
out-of-sample RV predictions (\cvrmsrvsscalpelstenoneoffour, \cvrmsrvsscalpelstentwooffour, \cvrmsrvsscalpelstenthreeoffour and \cvrmsrvsscalpelstenfouroffour \mps when using 10 feature vectors) and 
reaching its plateau reliably with fewer feature vectors.  
Based on comparing the predictive performance from analyzing the full data set and subsets for cross validation purposes, it appears that Scalpels's performance does degrade somewhat as the number of observations in the training sample decreases.
However, once Scalpels uses at least ten features, the RMS of Scalpels's out-of-sample cleaned RVs for the two runs become comparable (\cvrmsrvsscalpelstenpost~ \mps~ and \cvrmsrvsscalpelstenpre~ \mps).
This represents a substantial improvement in the level of stellar variability mitigation compared to all the other methods considered here and to previous published analyses (for  actual stellar observations as opposed to simulated data).

\section{Discussion}
\label{sec:discuss}
Here we discuss implications of our findings for NEID, mitigating stellar variability, and the design of future EPRV instruments and surveys. 

\subsection{RMS of Daily Cleaned RVs}
\label{sec:discuss_working_well}
The measured RMS of daily-averaged  RVs before attempting to clean them of stellar variability is \stdmeandailyrv~ \mps.
This greatly exceeds the photon noise for each daily binned spectrum (varies with the number of observations included, $\simeq$0.02--0.07 \mps) and the NEID instrumental error budget (0.27 \mps).
Decorrelating the daily RVs with any of the data-driven, wavelength-domain stellar variability mitigation strategies considered yielded a significant reduction in the RMS RV (resulting in an RV RMS of \rmsrvsscalpels -- \rmsrvsfiesta~ \mps~ depending on algorithm).  
Each data-driven algorithm we considered performed better than decorrelating against classical spectroscopic indicators or chromatic RV information.
Of particular note, Scalpels performed particularly well, 
reducing the RV RMS of daily cleaned data to \rmsrvsscalpels~ \mps~ and
achieving an out-of-sample RV prediction accuracy of \cvrmsrvsscalpelstenthreeoffour -- \cvrmsrvsscalpelstentwooffour \mps~ when training of approximately three quarters of the data.

As the NEID solar telescope continues to collect more data, we'll be able to reassess how much of the difference between these and the \rmsrvsscalpels~ \mps~ when applied to the full dataset is due to the reduction in the size of the training set and how much is due to out-of-sample predictions being harder than within sample predictions.   

\subsection{Effect of limiting integration time}
\label{sec:discuss_1hr}
\begin{figure}
    \centering
    \includegraphics[width=0.48\linewidth]{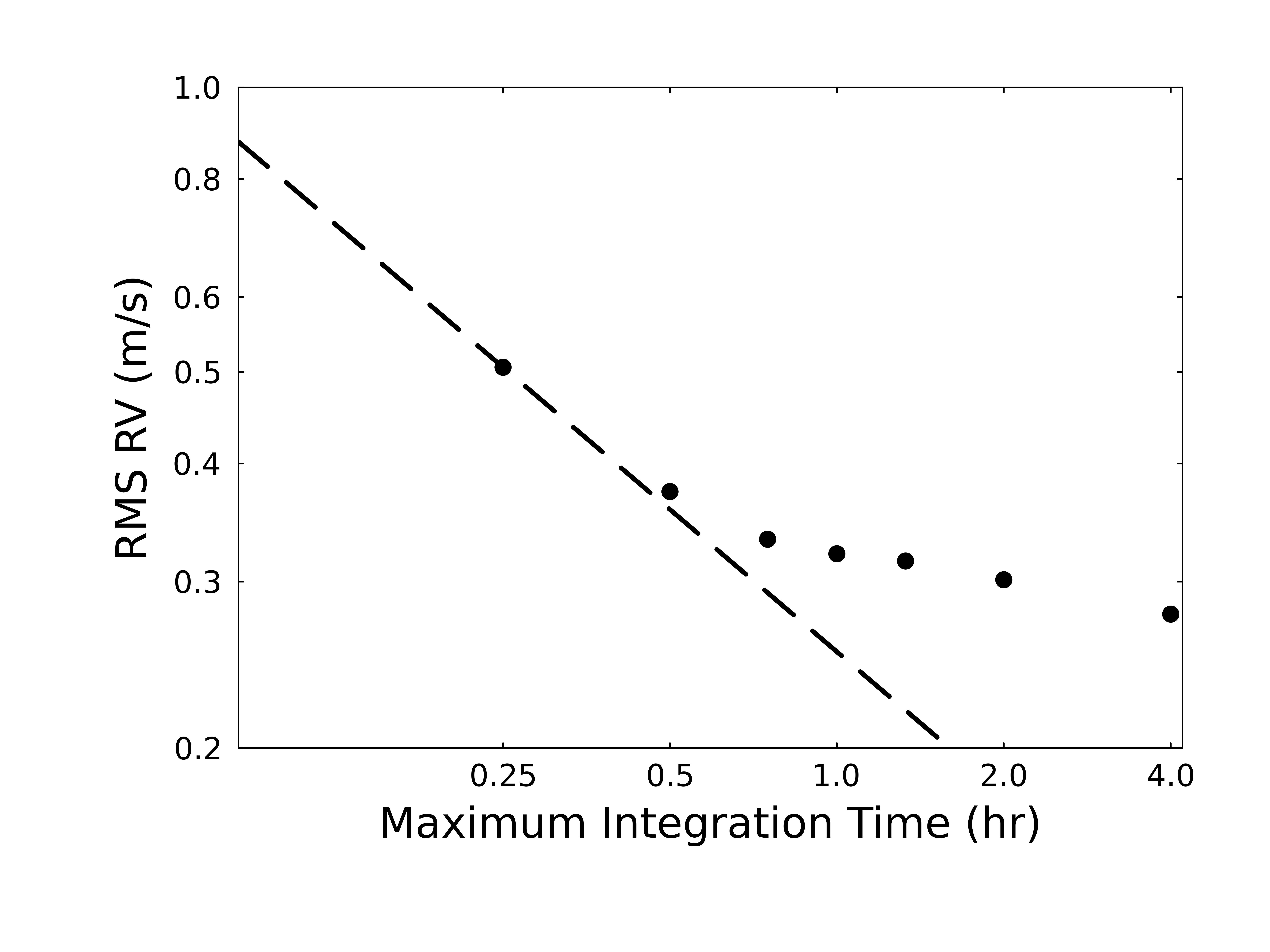}
    \caption{RMS of daily mean cleaned RV as a function of the maximum integration time allowed per day.  Daily average RVs are computed for days with at least fifteen binned observations (each bin consisting of five observations).  The number of binned observations included in each daily average CCF was limited, so the total integration time did not exceed the duration on the x-axis.  For days with more observations than needed, binned observations with the lowest airmass were selected.  Daily averaged CCFs were cleaned with Scalpels using 10 feature vectors.  The dashed line shows a $-\frac{1}{2}$ power law with arbitrary normalization.  \label{fig:rms_vs_inttime}}
    
\end{figure}

Given the success of Scalpels in cleaning daily averaged CCFs, one naturally wonders whether the excellent performance was only possible due to combining hundreds of spectra to obtain a very high effective signal-to-noise. 
Therefore, we repeated our Scalpels analysis of daily averaged spectra, but computing daily averaged CCFs that used no more than one hour of integration time each (see Figure \ref{fig:rms_vs_inttime}).  
Limiting the integrating time per day to no more than one hour increased
the RMS of the cleaned daily RVs by only $\simeq~16\%$ relative to our baseline analysis. 
In cross validation tests, the increase in RMS of daily cleaned RV based on just one hour of observations per day was similar, resulting in an RMS of $\simeq 0.4$ \mps.  

The RMS of cleaned RVs increases slowly as the maximum integration time decreases from four to $\simeq1$ hour, and then increases more rapidly for maximum integration times of less than 45 minutes.  
For comparison, the typical photon noise for measuring a pure Doppler shift using one hour of integration time is $\simeq0.045$ \mps.  
While Scalpels is  removing the impact of stellar variability to the $\simeq 0.3$ or 0.4 \mps~ level, doing so requires precisely characterizing the CCF shape and thus much higher signal-to-noise to than if the solar spectrum were constant and errors were dominated by photon noise.  

For Sun-like stars where magnetic activity is effectively suppressed using techniques such as Scalpels as shown here for the Sun, granulation and supergranulation are expected to become limiting factors for the accuracy of EPRV mass measurements \citep{Luhn2023CorrelatedNoise,PalumboGrassII}.  
Therefore, An additional potential factor in how well Scalpels can clean RV is the wall time of solar observations significantly exceeds the integration time.  
Solar oscillations, a complex beat pattern of stochastically excited p-modes, granulation, super granulation, and potentially other processes contribute to the observed solar RV variability on timescales of minutes to days \citep{EPRVWGReport}.  
A portion of the improvement as a function of integration time may be due to the greater wall time more effectively averaging over these processses.  
For night-time observations, the duty cycle is higher than for solar observations ($\simeq66\%$).
Therefore, night-time observations are expected to be more effective in averaging over oscillations than NEID solar observations.  
However, once granulation and/or supergranulation become significant sources of RV contamination, it may be important to make multiple visits to a target star within a night to spread observations over a greater wall time.

\subsection{Consistency of ML-based Detrending}
\label{sec:consistent_ml_detrend}
\begin{figure}
    \centering
    \includegraphics[width=0.48\linewidth]{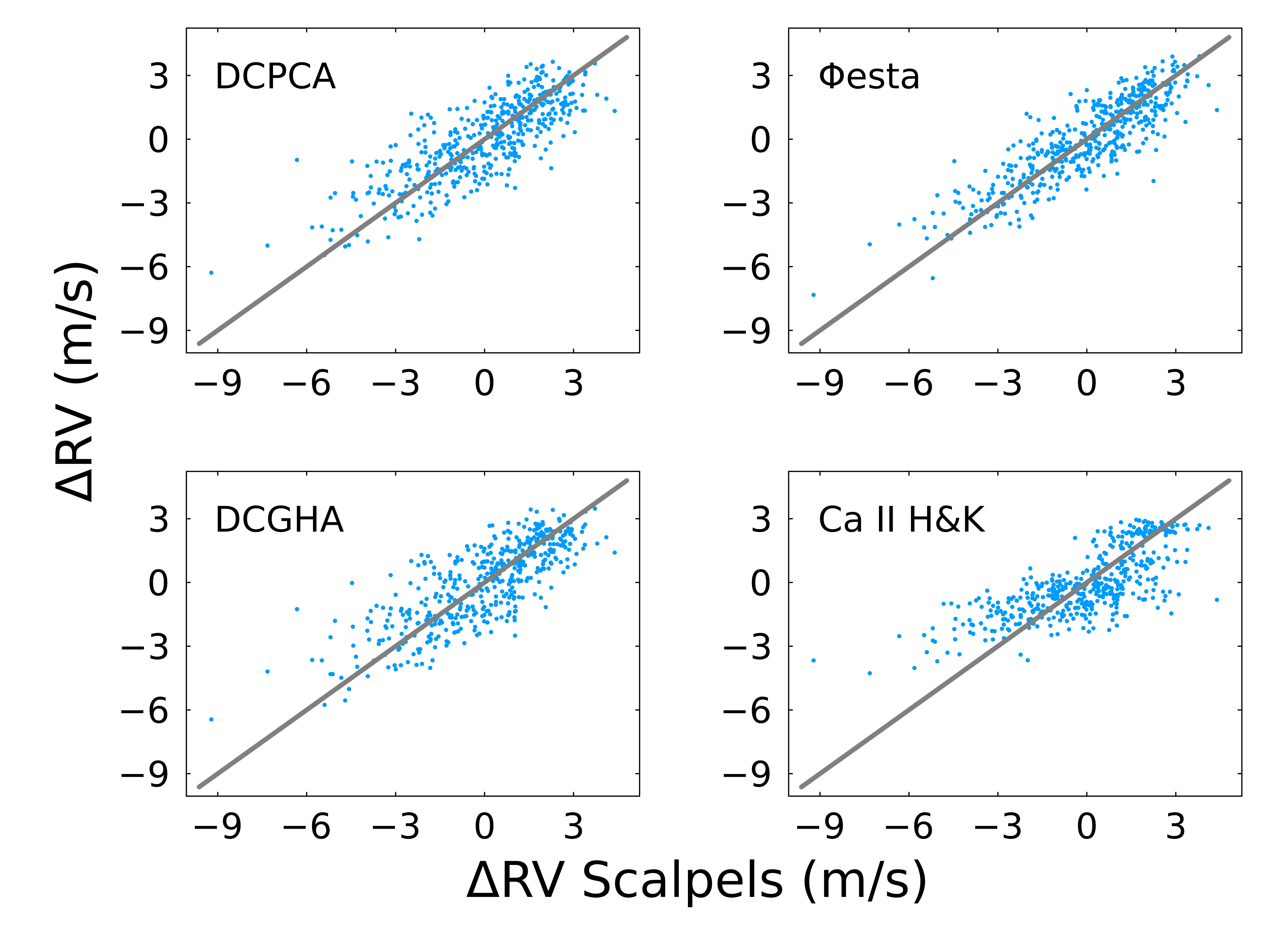}
    \caption{Comparison of the correction applied to daily-averaged RVs when detrending using different stellar variability indicators.  In each panel, the x-axis is the correction applied by Scalpels.  The y-axis is the correction is applied based on trending against scores from DCPCA (top left), $\Delta~RV_k$'s from FIESTA (top right), scores from projecting the CCF$_{\mathrm{shape}}$ onto a Gauss-Hermite basis (bottom left), and the classical Ca II H\&K activity indicator.\label{fig:delta_rv_vs_delta_rv}}
    
\end{figure}

Figure \ref{fig:delta_rv_vs_delta_rv} compares the RV corrections suggested by different methods.
We find that the corrections suggested by each algorithm considered are highly correlated with each other, in contrast to the findings of \citet{ZhaoEsspTwoComparison}.
This could be due to a variety of factors.
First, \citet{ZhaoEsspTwoComparison} included more algorithms that took a more diverse set of approaches to cleaning RVs.  For example, this study does not attempt to use any time information, while several methods presented in \citet{ZhaoEsspTwoComparison} did.
Second, since many teams performed their own prepossessing of data, the analyses in \citet{ZhaoEsspTwoComparison} were less homogeneous than in this study.
Third, this study focused solely on Solar observations and the daily binned observations are at higher signal-to-noise than the datasets from \citep{ZhaoEsspData}.
While each of these factors is likely to have contributed to the different findings, further research will be need to quantify how much is due to each factor.

\subsection{Comparison to NEID Error Budget}
\label{sec:compare_rms_to_budget}
After cleaning by Scalpels, the RMS of daily averaged RVs over the NEID solar dataset (\rmsrvsscalpels~ \mps) is significantly larger than the photon noise limit, but comparable to the 0.27 \mps\ expected based on NEID instrumental error budget \citep{NEID_budget}. While a formal error budget for the NEID solar telescope was not developed at the same level of rigour, clearly the solar telescope is also performing well, after applying the filters described in this paper. 
This provides empirical evidence that the NEID instrument and pipeline are approaching their design goals (at least for bright G-star targets after applying corrections for known pipeline issues described in \S\ref{sec:filter} and \S\ref{sec:daily_slope}).  
Further research will be needed to identify those terms in the error budget that are most significant after cleaning data of solar variability.  
The largest four terms in NEID's error budget are: software algorithms for data reduction, software algorithms for the calibration process, micro-telluric contamination, and sky fiber subtraction, each of which is budgeted 0.1 \mps.  
Our results in \S\ref{sec:daily_slope} suggest that the algorithms for data reduction and calibration in NEID DRP v1.3 have room for further improvement.
When we performed multiple analyses with different line list that were more/less aggressive in avoiding telluric contamination, we did see small, but potentially interesting changes in the results, suggesting that further research in avoiding telluric contamination may yield additional improvements. 
The sky fiber subtraction is not used for solar observations, so the solar observations are not a diagnostic for that term during night time observations.  

\subsection{Comparison to HARPS-N Solar RVs}
\label{sec:compare_harpsn}
The \rmsrvsscalpels~ \mps~ RMS of NEID's daily averaged RVs after cleaning by Scalpels is significantly less than the 1.25 \mps~ obtained by applying Scalpel to HARPS-N solar observations \citep{CollierCameronScalpels}.
This is particularly impressive, since the sun was significantly more active during much of the NEID observations than during the HARPS-N observations analyzed in \citet{CollierCameronScalpels}.
The explanation for this difference is unclear.  

One candidate explanation is NEID's improved environmental stability. 
Interestingly, the application of Scalpels to HARPS-N solar observations benefited from using 13 feature vectors, more than needed to achieve a much lower RMS of daily RVs with  NEID solar observations.  
\citet{CollierCameronScalpels} linked multiple feature vectors to instrumental interventions that likely led to changes in the instrument LSF.  
In contrast, inspecting the Scalpels scores for NEID solar observations does not reveal any feature that can be clearly mapped to changes in the instrument before and after the Contreras fire.
The smaller number of feature vectors needed here suggests that NEID's improvement environmental stability is paying off in terms of improved LSF stability and ability to attribute observed line shape variation to astrophysical variability.
We interpret this as evidence that NEID's improved LSF stability may contribute to better removal of spurious RVs due to stellar variability.  

Another potential explanation for the difference in results when comparing NEID and HARPS-N solar observations is NEID's use of both the pyrheliometer and exposure meter to reject observations affected by poor atmospheric conditions.  
Therefore, we recommending including/adding a pyrheliometer to other solar telescopes.
Yet another possibility is that it is more effective to train data-driven algorithms on datasets that include significant solar variability, and most previous studies using HARPS-N solar data have focused on a range of data when the Sun was relatively inactive.  
This explanation could be tested by performing updated analyses of HARPS-N solar observations that include a wider range of solar activity levels.

\subsection{Prospects for EPRV detection of Earth twins}
\label{sec:prospects_earths}
\citet{CollierCameronScalpels} applied Scalpels to synthetic a dataset based on 853 days of HARPS-N solar observations, but with Doppler signature of multiple hypothetical low-mass planets added.  
They found that Scalpels would allow for accurate mass measurements of low-mass planets, with RV amplitude uncertainties of $\simeq$0.066~\mps~.  
Scaling their results by the ratio of the RMS of cleaned daily RVs from NEID to the RMS of cleaned daily RVs from HARPS-N would yield an expected uncertainty on the Doppler amplitude of 0.015~ \mps, sufficient to obtain a ``6-$\sigma$'' detection for an Earth twin.  
Thus, it appears we are now on the precipice of having the technology to detect true Earth twins with modern EPRV spectrographs (and generous observing time allocations).  
\S\ref{sec:future_beyond_sun} discusses some of the key remaining challenges.

\subsection{Summary of key findings}
\label{sec:summary}
Our findings show that the NEID solar telescope is performing extremely well, measuring solar RVs accurate to \medianrmsbinnedrv~ \mps, once binning over p-mode pulsations. 
This and the measured RMS of daily mean RV observations (\stdmeandailyrv ~\mps) are comparable to recent results from other solar telescopes such as those operating at HARPS-N, EXPRES, and HARPS \citep{ZhaoEsspThree}.  
We find that postprocessing the NEID daily averaged CCFs with one of several data-driven techniques can yield significant reduction in the accuracy of daily averaged solar RVs. 
The Scalpels algorithm stands out for its excellent performance in mitigating solar variability, reaching an RMS of daily averaged spectra of just \rmsrvsscalpels~ \mps.  
Even a strenuous cross validation test using only half of the available data for training yielded out-of-sample RMS of daily RV predictions of \cvrmsrvsscalpelstenpost -- \cvrmsrvsscalpelstenpre~ \mps.   
Limiting the integration time per day to 1 hour resulted in only a modest increase in the RMS of RVs after cleaning by Scalpels.  
Based on scaling the results from a previous study that jointly fit Scalpels with planetary scaling, it appears that detecting and characterizing the mass of an Earth-twin may now be within reach of existing EPRV instruments, if
provided sufficient observing time.

This motivates further effort devoted to refining pipelines and data analysis techniques in hopes of reducing the amount of observing time required and to allow EPRV facilities to search as many nearby stars as practical.  
To support further research and future comparisons to other stellar variability mitigation strategies, we provide a large public data set that has been consistently processed and cleaned of known instrumental effects. 
By applying algorithms to the exact same observations, the results of subsequent analyses can be directly compared to the results presented here.

\subsection{Future research opportunities}
\label{sec:future}
While our results are very encouraging for the prospects of NEID and other EPRV spectrographs to discover and characterize rocky planets in the habitable zone of Sun-like stars, there are still several areas ripe for further research. 

\subsubsection{Calibration and pipeline improvements}
\label{sec:future_pipeline}
First, we find opportunities for further improving the NEID wavelength calibration and drift model, both within each day and across days.  
It seems likely that improvements in the drift model could lead to Scalpels and other methods becoming even more effective at removing spurious RVs due to variability.    
This finding demonstrates that one of the powerful benefits of making solar observations is the ability to stress test an instrument and its data reduction pipeline, so as to motivate future improvements in data process, operations, and/or hardware.  
Similar analyses of solar observations from other EPRV instruments would likely be helpful in characterizing their performance, prioritizing issues to be improved, and enabling more precise cross-instrument comparisons.

\subsubsection{Extending to night-time observations of other stars}
\label{sec:future_beyond_sun}
Second, our analysis was restricted to solar observations that offer several advantages over night time observations.
It is practical to obtain hundreds of solar observations every clear day with a favorable observing cadence since solar observations do not impinge on nighttime observing programs.  
While each individual exposure is not particularly high signal-to-noise, combining information from several hours of observations effectively provides access to unusually high signal-to-noise data.  
Additionally, combining solar data taken over hours helps to suppress stellar variability due to oscillations and granulation.  
Additional research is needed to understand how the performance of algorithms such as Scalpels degrades as the number of observations, signal-to-noise, and integration time observations degrades. 
A preliminary exploration indicates the performance is only mildly degraded ($\simeq~16$\%) if the total integration time per day is reduced to 1 hour.  
Further research will be needed to explore the interaction of integration time and wall time (for marginalizing over granulation), as well as how these vary with stellar type.

Third, night time observations have some additional sources of noise or systematics that do not affect solar observations significantly.  
For example, solar observations due not need to worry about removing background contamination via a sky fiber.  
As another example, the throw of barycentric correction for solar observations is $\sim 1$\kmps, while other targets can have up to $\simeq 30$\kmps~ throw.  
Here we used a line list which was cleaned to avoid lines that would suffer significant contamination even if the barycentric corrections reached a 30 \kmps.  
However, further research will be needed to characterize the performance of stellar variability mitigation algorithms in the presence of both stellar and telluric variability.

Finally, additional research is needed to understand how these algorithms perform on other stars and to what extent solar data can be used for training and testing stellar variability mitigation techniques for other stellar types.  
This is particularly important for EPRV to play a supporting role in discovery and mass characterization of potentially Earth-like planets for future large direct imaging missions (e.g., Habitable Worlds Observatory).

\subsubsection{Combining wavelength \& time-domain information}
\label{sec:future_comb_wave_time}
Stellar mitigation strategies can be divided into three broad categories:  
(1) wavelength-domain strategies that use changes in each stellar spectrum individually to distinguish true Doppler shifts from stellar variability, 
(2) time-domain strategies that use the known shape true planetary signals to separate true Doppler shifts from stellar variability, and 
(3) hybrid strategies that analyze spectroscopic timeseries jointly, so as to combine wavelength and time-domain information.   
In \S\ref{sec:eval_mitigation}, we explored several wavelength-domain strategies that make use of classical spectroscopic indicators, the CCF bisectors, or machine learning approaches using input features computed from the CCF.  
In this study, we have considered mitigating the impacts of solar variability on derived RVs using only the instantaneous spectrum, rather than incorporating temporal information.  
Wavelength-domain approaches has the significant benefit of improving the interpretability and explainability of results relative to hybrid strategies.  
We recommend future studies to consider time-domain and hybrid strategies.
In particular, previous studies have found that some stellar activity indicators appear to have a lag, i.e., reach their local maximum a few days before/after the spurious RV reaches a local extrema.  
Complementary techniques such as Gaussian process regression using a latent process and its derivatives \citep[e.g.,][]{GilbertsonGlom}, a Gaussian process with a non-standard kernel that explicitly incorporates a lag, or a convolutional neural network may be capable of making better use of the stellar variability indicators considered here. 
We defer making use of temporal information for future studies.

\begin{acknowledgements}
We thank Alex Wise for his contributions to developing \verb|RvSpectML|, espsecially the \verb|RvLineList.jl| package.
This work was partially supported by the  Heising-Simons Foundation award \#2019-1177 (E.B.F., S.M., J.T.W, J.Z) and NASA Grant \# 80NSSC21K1035 (E.B.F., S.M., R.C.T.).
This work was supported in part by a grant from the Simons Foundation/SFARI (675601, E.B.F.).
.
%
S.M.\ is the NEID Principal Investigator.
P.R.\ serves as NEID Instrument Team Project Scientist.  
%
%
%
%
%
The Center for Exoplanets and Habitable Worlds and the Penn State Extraterrestrial Intelligence Center are supported by Penn State and its Eberly College of Science.
Computations for this research were performed on the Pennsylvania State University’s Institute for Computational and Data Sciences’ Roar supercomputer.


Part of this work was performed at the Jet Propulsion Laboratory, California Institute of Technology, sponsored by the United States Government under the Prime Contract 80NM0018D0004 between Caltech and NASA. 

Based 
 on observations at Kitt Peak National Observatory, NSF’s NOIRLab 
 , managed by the Association of Universities for Research in Astronomy (AURA) under a cooperative agreement with the National Science Foundation. The authors are honored to be permitted to conduct astronomical research on Iolkam Du\'ag (Kitt Peak), a mountain with particular significance to the Tohono O\'odham.
 
 Data presented herein were obtained at the WIYN Observatory from telescope time allocated to NN-EXPLORE through the scientific partnership of the National Aeronautics and Space Administration, the National Science Foundation, and the National Optical Astronomy Observatory.
We thank the NEID Queue Observers and WIYN Observing Associates for their skillful execution of our NEID observations.
Deepest gratitude to Zade Arnold, Joe Davis, Michelle Edwards, John Ehret, Tina Juan, Brian Pisarek, Aaron Rowe, Fred Wortman, the Eastern Area Incident Management Team, and all of the firefighters and air support crew who fought the recent Contreras fire. Against great odds, you saved Kitt Peak National Observatory.


This research has made use of 
NASA's Astrophysics Data System Bibliographic Services.




%

\end{acknowledgements}

\facilities{WIYN (NEID)}

\software{
EchelleCCFs.jl v0.2.9 (\url{https://github.com/RvSpectML/EchelleInstruments.jl}), 
EchelleInstruments.jl v0.2.11 (\url{https://github.com/RvSpectML/EchelleInstruments.jl}), 
NeidArchive.jl v0.1.3 (\url{https://github.com/RvSpectML/NeidSolarScripts.jl}), 
NeidSolarScripts.jl v0.1.6 (\url{https://github.com/RvSpectML/NeidSolarScripts.jl}), 
RvLineList.jl (\url{https://github.com/RvSpectML/RvLineList.jl}), 
RvSpectML.jl v0.2.8 (\url{https://github.com/RvSpectML/RvSpectML.jl}),
RvSpectMLBase.jl v0.2.3 (\url{https://github.com/RvSpectML/RvSpectMLBase.jl}), 
SunAsAStar.jl v0.1.2 (\url{https://github.com/RvSpectML/SunAsAStar.jl}) 
}

\bibliography{NEIDreferences,solar_telescopes,stellar_variability,specific_to_this_paper}

\end{document}